\documentclass[usenatbib]{mn2e}
\usepackage{graphicx}
\usepackage{savesym}
\usepackage{amsmath}
\savesymbol{iint}
\usepackage{txfonts}

\usepackage{subfigure}
\usepackage{epsfig}
\usepackage{url}
\def\simless{\mathbin{\lower 3pt\hbox
{$\rlap{\raise 5pt\hbox{$\char'074$}}\mathchar"7218$}}}   
\def\simmore{\mathbin{\lower 3pt\hbox
{$\rlap{\raise 5pt\hbox{$\char'076$}}\mathchar"7218$}}}   
\def\Msun{{\rm M}_\odot}                                       
\newcommand{\be}{\begin{equation}}
\newcommand{\ee}{\end{equation}}
\newcommand       \bea          {\begin{eqnarray}}
\newcommand       \eea          {\end{eqnarray}}
\newcommand       \apj          {ApJ}
\newcommand       \apjl         {ApJL}
\newcommand       \aap          {A\&A}
\newcommand       \nat          {Nature}
\newcommand       \mnras        {MNRAS}

\newcommand       \aj      {AJ}
\newcommand       \prd      {Phys.~Rev.~D.~}

\newcommand       \araa      {ARA\&A}

\newcommand      \apjs {ApJ Supplements}

\def\simlt{\mathrel{\hbox{\rlap{\hbox{\lower4pt\hbox{$\sim$}}}\hbox{$<$}}}}
\def\simgt{\mathrel{\hbox{\rlap{\hbox{\lower4pt\hbox{$\sim$}}}\hbox{$>$}}}}
\def\lesssim{\mathrel{\hbox{\rlap{\hbox{\lower4pt\hbox{$\sim$}}}\hbox{$<$}}}}

\def\gtrsim{\mathrel{\hbox{\rlap{\hbox{\lower4pt\hbox{$\sim$}}}\hbox{$>$}}}}

\title[The TDE Delay Time Distribution]{The Delay Time Distribution of Tidal Disruption Flares}

\author[N.C. Stone et al.]{Nicholas C.~Stone$\thanks{E-mail: nstone@phys.columbia.edu}^{1}$, Aleksey Generozov$^1$, Eugene Vasiliev$^{2,3}$, Brian D. Metzger$^1$ \\
$^1$Columbia Astrophysics Laboratory, Columbia University, New York, NY, 10027, USA \\
$^2$Rudolf Peierls Centre for Theoretical Physics, Oxford, OX1 3NP, UK\\
$^3$Lebedev Physical Institute, Moscow, 119991, Russia\\ 
$^\star$Einstein Fellow; nstone@phys.columbia.edu}

\begin{document}
\date{Received / Accepted}
\pagerange{\pageref{firstpage}--\pageref{lastpage}} \pubyear{2014}

\maketitle

\label{firstpage}

\begin{abstract}
Recent observations suggest that stellar tidal disruption events (TDE) are strongly overrepresented in rare, post-starburst galaxies.  Several dynamical mechanisms have been proposed to elevate their TDE rates, ranging from central stellar overdensities to the presence of supermassive black hole (SMBH) binaries.  Another such mechanism, introduced here, is a radial velocity anisotropy in the nuclear star cluster produced during the starburst, which temporarily enhances the stellar flux into the loss cone of a solitary SMBH.  These, and other, dynamical hypotheses can be disentangled by comparing observations to theoretical predictions for the TDE {\it delay time distribution} (DTD).  We show that SMBH binaries are a less plausible solution for the post-starburst preference, as they predict an unrealistically top-heavy distribution of primary SMBH masses, and can only reproduce the observed DTD with extensive fine-tuning.  The overdensity hypothesis produces a reasonable match to the observed DTD (based on the limited data currently available), provided that the initial stellar density profile created during the starburst, $\rho(r)$, is exceptional in both steepness and normalization.  In particular, explaining the post-starburst preference requires $\rho \propto r^{-\gamma}$ with $\gamma \gtrsim 2.5$, i.e. much steeper than the classic Bahcall-Wolf equilibrium profile of $\gamma = 7/4$.  For ``ultrasteep'' density cusps ($\gamma \ge 9/4$), we show that the TDE rate decays with time measured since the starburst as $\dot{N} \propto t^{-(4\gamma-9)/(2\gamma-3)} / \ln t$.  Radial anisotropies also represent a promising explanation, provided that initial anisotropy parameters of $\beta_0 \approx 0.5$ are sustainable against the radial orbit instability.  TDE rates in initially anisotropic cusps will decay roughly as $\dot{N} \propto t^{-\beta_0}$.  As the sample of TDEs with well-studied host galaxies grows, the DTD will become a powerful tool for constraining the exceptional dynamical properties of post-starburst galactic nuclei.
\end{abstract}

\section{Introduction}
\label{sec:intro}
Stellar tidal disruption events (TDEs) have long been seen as powerful tools for studying the demography of quiescent supermassive black holes (SMBHs).  The death of a star during a strong tidal encounter with an SMBH \citep{Hills75} is expected to produce a luminous, multiwavelength flare \citep{Rees88}.  Many such flares have now been detected through thermal emission in optical \citep{vanVelzen+11, Gezari+12, Chornock+14, Holoien+14, Blagorodnova+17, Hung+17}, ultraviolet \citep{Gezari+06, Gezari+08}, and soft X-ray \citep[see e.g.][]{Komossa15, Auchettl+17} wavelengths, while others have been seen through nonthermal emission in the radio \citep{Zauderer+11, vanVelzen+16} or hard X-ray \citep{Bloom+11, Cenko+12, Brown+15}.  The observed sample of TDEs is expected to grow from dozens to thousands in the near future \citep{vanVelzen+11, Khabibullin+14}.  This large future sample carries great promise for studying SMBH demography, although there are many open theoretical questions that must be resolved before TDE light curves can be translated into SMBH mass \citep{Lodato+09, GuillochonRamirezRuiz13, Hayasaki+13, Piran+15} or spin \citep{StoneLoeb12, Hayasaki+16, Franchini+16} measurements.

While the prospects for using TDE light curves to study SMBH demography lie in the future, the statistical properties of our current TDE sample {\it have already} revealed unexpected dynamical processes in distant galactic nuclei.  \citet{Arcavi+14} were the first to notice the peculiar host galaxy preference of observed optically-selected TDEs: in the Palomar Transient Factory sample of three strong TDE candidates, two were found in E+A galaxies, a rare, post-starburst galaxy type that makes up $\approx 0.2\%$ of all galaxies in the low-redshift universe.  E+A galaxies exhibit strong H$\delta$ absorption features indicative of many young A stars, but little to no H$\alpha$ emission line strength, signaling an absence of ongoing star formation.  Subsequent observations of TDE host galaxies also found that an order unity fraction of these flares inhabit this exceedingly uncommon galaxy subtype.  

\citet{French+16, French+17} conducted the first thorough statistical examination of TDE host galaxy properties, finding that both classical E+A galaxies and more weakly post-starburst galaxies were strongly overrepresented in their host sample.  This second host galaxy subpopulation exhibits a similar dearth of ongoing star formation but weaker H$\delta$ absorption features.  Its combination of emission and absorption properties is found in $\approx 2\%$ of low redshift galaxies, and is consistent with an older post-starburst galaxy, or alternatively a young post-starburst galaxy that underwent a weaker episode of star formation (\citealt{French+17} find some degeneracy between these two explanations of strong H$\delta$ features).  In a sample of 8 TDE hosts, classical E+As and weaker post-starburst galaxies were overrepresented by factors of $\mathcal{R}=190^{+115}_{-100}$ and $\mathcal{R}=33^{+7}_{-11}$, respectively.  Qualitatively similar results were found by a more recent analysis of a larger sample of optically and X-ray bright tidal disruption flare hosts \citep{Graur+17}, though with a somewhat reduced overrepresentation of post-starburst galaxies among TDE hosts (using the same cuts on H$\delta$ absorption strength, the rare host overrepresentation fell to factors of $\mathcal{R}=36^{+22}_{-18}$ and $\mathcal{R}=18^{+8}_{-6}$; see also \citealt{LawSmith+17} for complementary analysis).  TDE hosts clearly exhibit a strong post-starburst preference (PSP), though the exact magnitude of this preference remains an area of active research.

This finding is even more puzzling when juxtaposed against a broader TDE rate discrepancy \citep{StoneMetzger16}: observational estimates, although limited by small sample sizes and selection effects that are difficult to quantify, often find per-galaxy TDE rates $\dot{N} \sim 1-2\times 10^{-5}~{\rm gal}^{-1}~{\rm yr}^{-1}$ \citep{Donley+02, vanVelzenFarrar14}.  These are at least an order of magnitude below conservative theoretical estimates for TDE rates (set by two-body relaxation) in realistic galactic nuclei, which are typically $\dot{N} \gtrsim 10^{-4} ~{\rm gal}^{-1}~{\rm yr}^{-1}$ \citep{WangMerritt04, StoneMetzger16}, although we note that some observed TDE samples find rates up to $\dot{N} \sim 1-2\times10^{-4}~{\rm gal}^{-1}~{\rm yr}^{-1}$ \citep{Esquej+08, Holoien+16}.  Although dynamical mechanisms to suppress TDE rates have been proposed, they seem unlikely to work in practice \citep{LezhninVasiliev15, LezhninVasiliev16}.  The resolution of this rate discrepancy (which is worsened by the PSP) may be a very broad TDE luminosity function (e.g.~\citealt{Blagorodnova+17}), of which we have so far only seen the bright end.  Indeed, recent modeling of the optical TDE luminosity function suggests an observational rate $\dot{N} \approx 1\times 10^{-4}~{\rm gal}^{-1}~{\rm yr}^{-1}$ \citep{vanVelzen17}.  The broader rate discrepancy is not yet a solved problem, but in this paper we make the reasonable assumption that the TDE luminosity function is the same in post-starburst and normal galaxies, and search for dynamical explanations for the PSP\footnote{Non-dynamical effects - e.g. preferential nuclear obscuration in normal galaxies, that is absent in post-starburst galaxies - could contribute to both the PSP and the overall TDE rate discrepancy, but in this paper we focus on dynamical hypotheses for intrinsically elevated rates of disruption in post-starburst galaxies.}. 

What dynamical mechanisms could be elevating the TDE rate so dramatically following a starburst?  \citet{Arcavi+14} proposed that, as many starbursts are triggered by galaxy mergers, the PSP may reflect a population of SMBH binaries (SMBHBs) which, as they harden, pass briefly through a stage where TDE rates are increased by many orders of magnitude.  \citet{StoneMetzger16} suggested that if the starburst is preferentially concentrated in the galactic nucleus, then a strong stellar overdensity may be formed, which would increase TDE rates by decreasing the two-body relaxation timescale.  This overdensity hypothesis has received tentative observational support from observations and dynamical modeling of one of the nearest E+A galaxies, NGC 3156 \citep{StonevanVelzen16}; likewise, broader population studies find that per-galaxy TDE rates are correlated with both the slope \citep{LawSmith+17} and the normalization \citep{Graur+17} of stellar light profiles on $\sim {\rm kpc}$ scales (though we caution that the TDE rate is set by stellar dynamics on $\sim {\rm pc}$ scales).  Other possibilities exist as well: rates could be enhanced due to non-conservation of orbital angular momentum in a triaxial potential created by starburst or galaxy merger \citep{MagorrianTremaine99, MerrittPoon04}, or due to secular dynamics in eccentric nuclear stellar disks \citep{Madigan+17}.

We propose that the several extant hypotheses for the PSP, as well as many more yet to be suggested by theorists, can be disentangled by empirical construction of a TDE {\it delay time distribution} (DTD).  Different dynamical mechanisms for enhancing TDE rates in a starburst will attenuate over time, as the galaxy ages and its nuclear properties evolve to resemble those of more typical galaxies.  But different dynamical solutions to the PSP will make different predictions for exactly how TDE rates decline with time since the starburst, and these predictions can be tested against observations.  The results of \citet{French+16}, \citet{French+17}, and \citet{Graur+17} represent the first empirical TDE DTDs, although they are limited by small number statistics.  In the near future, the {\it Zwicky Transient Facility} (ZTF) and {\it Large Synoptic Survey Telescope} (LSST) will discover tens to thousands of TDEs per year, respectively, and empirical DTDs may definitively test theoretical explanations for the PSP.

In this paper, we estimate rates and, when possible, DTDs for three different dynamical hypotheses that aim to explain the post-starbust preference.  In \S \ref{sec:overdense}, we review the basis for the overdensity hypothesis and make simple estimates for the time evolution of TDE rates in overdense galactic nuclei.  In \S \ref{sec:aniso}, we introduce a new dynamical explanation for the PSP, one predicated on velocity-space anisotropies biased toward radial orbits; we also estimate the time evolution of TDE rates in this scenario.  In \S \ref{sec:SMBHB}, we review the SMBHB hypothesis and, again, estimate how quickly an elevated TDE rate decays against time since starburst.  In \S \ref{sec:DTD} we combine our analysis of the prior three scenarios and make a first comparison to the observed DTD of tidal flares, and in \S \ref{sec:conclusions} we summarize this work and offer thoughts on its future extensions.

\section{Overdense Stellar Systems}
\label{sec:overdense}

Empirically, the starbursts that create E+A galaxies preferentially concentrate star formation in their central regions.  Spectroscopic observations of E+As find strong radial color gradients, indicating that the youngest stars are concentrated in the galactic nucleus \citep{Pracy+12, Pracy+13}, though the resolution of these studies has historically been insufficient to provide detailed information on the SMBH influence radius.  Given the overrepresentation of nuclear regions in the star formation history of post-starburst galaxies, it is natural to speculate that central overdensities are responsible for the PSP: overdense galactic nuclei will possess short two-body relaxation times, and the rapid diffusion of stars through orbital phase space will result in a high rate of stellar tidal disruption.

Some support for this hypothesis was found by \citet{StonevanVelzen16}, who analyzed archival {\it Hubble Space Telescope} ({\it HST}) photometry of the nearby E+A galaxy NGC 3156, and found an unusually steep stellar density cusp which should produce an elevated TDE rate ($\dot{N} \approx 1 \times 10^{-3}~{\rm gal}^{-1} {\rm yr}^{-1}$) due to rapid two-body relaxation.  However, the small sample size of this study means that confirmation of the overdensity hypothesis will need to wait for future {\it HST} observations of other nearby post-starburst galaxies.

Enhanced TDE rates due to central stellar overdensities will decline over time, as two-body relaxation diffuses stars onto wider orbits, or as the dense stellar population drains into the SMBH.  To quantify this decline, it is useful to consider how two-body relaxation diffuses stars into the SMBH loss cone (LC),  the region of stellar phase space where specific orbital angular momentum $J$ is less than a critical value $J_{\rm LC} = \sqrt{2GM_\bullet R_{\rm t}}$.  Here we have defined the tidal radius $R_{\rm t}\equiv R_\star (M_\bullet /M_\star)^{1/3}$ as the radius interior to which a star of mass $M_\star$ and radius $R_\star$ will be torn apart by tides from a black hole of mass $M_\bullet$.  Near the SMBH, per-orbit perturbations (from two-body relaxation, or other processes) to specific angular momentum $\Delta J$ are weak, with $\Delta J / J_{\rm LC} \ll 1$; here the LC is devoid of stars and the rate of stellar tidal disruption is the rate at which stars diffuse into it.  Far from the SMBH, $\Delta J / J_{\rm LC} \gg 1$ and the LC is full; the per-star TDE rate here is, crudely, the fractional size of the LC in angular momentum space divided by a stellar orbital period.  The transition between these two regimes occurs at the critical radius $r_{\rm crit}$ where $\Delta J=J_{\rm LC}$ \citep{Merritt13}.

For spherically symmetrical galactic nuclei, time-dependent TDE rates can be found by numerically integrating the 2D (energy and angular momentum) Fokker-Planck equation \citep{CohnKulsrud78}, and quasi-steady state rates can be derived in a more semi-analytic way \citep{MagorrianTremaine99, WangMerritt04}.  We review some results from the quasi-steady state LC formalism in Appendix \ref{app:LC}, but the basic picture is as follows.  Starting from a 3D stellar mass density profile $\rho(r)$ and gravitational potential $\psi(r)$, we compute the stellar distribution function (DF) $f(\epsilon, J)$.  In general this is a two-integral DF, written in terms of $J$ and specific energy\footnote{In this paper we adopt the stellar dynamics convention where bound orbits have $\epsilon>0$.} $\epsilon$, but if velocities are isotropic it simplifies to $f(\epsilon)$.  Using the DF, one finds orbit-averaged diffusion coefficients $\bar{\mu}(\epsilon)$ for the radial orbits of interest, and from these one can calculate the flux of stars into the LC, $\mathcal{F} (\epsilon)$.  The total TDE rate is $\dot{N} = \int \mathcal{F}(\epsilon){\rm d}\epsilon$.  We will first perform this excercise for a fully analytic toy model of $\rho(r)$ and $f(\epsilon)$, and then repeat the procedure using time-dependent numerical solutions to the 1D (energy) Fokker-Planck equation.

\subsection{Analytic Evolution of Stellar Overdensities}
\label{sec:analytic}

Consider an idealized spherical profile of stars: $\rho(r) = \rho_{\rm infl} (r/r_{\rm infl})^{-\gamma}$, where the SMBH influence radius $r_{\rm infl}$ is defined as the radius that encloses the SMBH's own mass in stars, so $\rho_{\rm infl}=M_\bullet(3-\gamma)/(4\pi r_{\rm infl}^3)$.  For simplicity we will assume an isotropic distribution of velocities (but see \S \ref{sec:aniso} for the more general anisotropic case); this yields a stellar DF $f(\epsilon) \propto \epsilon^{\gamma-3/2}$.  In this model, there are two qualitatively different locations from which the bulk of the LC flux originates.  In typical, relatively shallow density cusps ($\gamma < 9/4$), most of the LC flux comes from radii $r\approx r_{\rm crit}$.  However, in much steeper density cusps ($\gamma > 9/4$), the LC flux actually {\it diverges} at small radii \citep{SyerUlmer99}; in reality, this indicates that the single power law assumption is not self-consistent, and most of the tidal disruptions will be sourced from whatever small radius corresponds to a break in the larger scale power law density profile.  For the remainder of this paper, we refer to $\rho(r)$ profiles with $\gamma>9/4$ as ``ultrasteep.''  One can imagine two ways in which centrally concentrated star formation could provide an elevated TDE rate:
\begin{itemize}
\item The formation of an ultrasteep cusp in a galactic nucleus with a relatively normal value of $r_{\rm infl}$.
\item The formation of a very dense stellar cusp with a more typical slope $\gamma$ but an atypically small value of $r_{\rm infl}$.  We refer to such a cusp as ``overconcentrated.''
\end{itemize}
Because stars are typically fed into the LC from the influence radius or smaller scales, both of these scenarios are compatible with starbursts that only increase the total galactic bulge mass by small amounts (see e.g. the burst mass fractions estimated among TDE hosts in \citealt{French+17}).  The latter scenario (low $r_{\rm infl}$) would require an order unity or greater increase in stellar mass within the pre-starburst influence radius, while the former (large $\gamma$) scenario requires even less total star formation.  These two possibilities are not mutually exclusive.

Almost all observed galactic nuclei possess $\gamma < 9/4$
\citep{Lauer+05}, but \citet{StonevanVelzen16} found that the nucleus
of NGC 3156, a nearby E+A galaxy, is at the very least a borderline case and may
indeed be ultrasteep; depending on the precise model employed for the
point spread function of {\it HST}, this galaxy has $2.15\le \gamma
\le 2.31$\footnote{NGC 3156 also possesses an unusually small influence radius, although measuring this quantity requires extrapolating the fitted power law slope beyond the {\it HST} resolution limit.}.  If we assume that most of the central stars formed
impulsively in a starburst, these (and less steep) central density
cusps will relax over time towards a steady state configuration.  For
a single-species stellar present day mass function (PDMF), this is the
well known Bahcall-Wolf cusp, with $\gamma=7/4$ \citep{BahcallWolf76}.
Multi-species PDMFs will exhibit more complicated density profiles
\citep{BahcallWolf77, Keshet+09}, but we defer an investigation of
this to future work.

So long as the starburst forms a density profile with central $\gamma > 3/2$, relaxation times decrease with decreasing radius $r$, and the profile relaxes towards a Bahcall-Wolf configuration from the inside-out.  Because we are interested primarily in the densest galactic nuclei, we limit ourselves to this regime\footnote{In principle, one could imagine a starburst producing a shallow power law ($\gamma<3/2$) with an almost discontinuous density jump to a high- but constant-density core.  This scenario would be of interest for the overdensity hypothesis, but would not be well approximated by this section, and should also be considered in future work.}, and approximate a time dependent density profile as the broken power law
\begin{equation}
\rho(r, t)=
\begin{cases}
 \rho_{\rm infl}(r/r_{\rm infl})^{-\gamma}, &r>r_{\rm BW}(t) \\
 \rho_{\rm infl}(r/r_{\rm BW})^{-7/4}(r_{\rm BW}/r_{\rm infl})^{-\gamma}, &r\le r_{\rm BW}(t), \label{eq:rhoBroken}
\end{cases}
\end{equation}
where we have defined a ``Bahcall-Wolf radius'' $r_{\rm BW}$ as the location where
the post-starburst age $t$ equals the local energy relaxation time
\citep{StonevanVelzen16}. In the Keplerian potential of the
SMBH, the relaxation time is
\begin{equation}
  t_{\rm r}(r)= \frac{\sigma^2(r)}{\langle \Delta (v_{\parallel})^2
    \rangle }  \label{eq:trx} = \frac{k(\gamma) (G M_\bullet)^{3/2} \langle m_\star \rangle}{G^2
    \langle m_\star^2 \rangle \rho_{\rm infl} r_{\rm
      infl}^\gamma \ln \Lambda} r^{\gamma-3/2}
\end{equation}
where $\sigma^2(r)=GM_\bullet r^{-1}/(1+\gamma)$ is the one-dimensional velocity dispersion, $\langle (v_\parallel )^2\rangle$ is a local diffusion coefficient defined in Appendix \ref{app:LC}, and the dimensionless constant $k(\gamma) \approx 0.05$ is defined in that appendix.  Here $\langle m_\star \rangle$ is the first moment of the stellar mass function and $\langle m_\star^2 \rangle$ is the second moment. The Bahcall-Wolf radius is thus
\begin{equation}
r_{\rm BW} = \left( \frac{k(\gamma) M_\bullet^{3/2} \langle m_\star \rangle}{G^{1/2} \langle m_\star^2 \rangle \rho_{\rm infl} r_{\rm infl}^\gamma t \ln \Lambda} \right)^{\frac{1}{3/2-\gamma}}, \label{eq:rBW}
\end{equation}
where we have used the Coulomb logarithm $\Lambda \approx 0.4 M_\bullet / \langle m_\star \rangle$.  An important caveat to this estimate is that in Eq. \ref{eq:trx}, we have defined the relaxation time in terms of a local diffusion coefficient.  Although this is a reasonable approximation in standard galactic nuclei, it is fundamentally nonlocal, orbit-averaged diffusion coefficients which control energy relaxation.  In cusps with $\gamma \ge 2$, orbit-averaged diffusion coefficients become strongly nonlocal.  While the basic timescale argument behind Eq. \ref{eq:rBW} remains correct, the simulations of \S \ref{sec:overdenseNumerical} allow us to calibrate a fitting formula that better matches Eq. \ref{eq:rBW} to numerical results: $k(\gamma) \approx 0.045(3-\gamma)$.

In this toy model, we can apply standard LC theory to extract the time
dependent TDE rate $\dot{N}$ (and therefore the DTD) from the density profile given by Eq. \ref{eq:rhoBroken}.  In the ultrasteep regime, the DTD can be approximated in closed form using an
analytic expression for LC flux $\mathcal{F}(\epsilon)$ from stars
deep inside the influence radius of the SMBH \citep[Appendix
A]{StoneMetzger16}.  In this regime, 
\begin{equation}
\dot{N}_\rho(t) \sim \epsilon_{\rm BW}(t)\mathcal{F}(\epsilon_{\rm BW}(t)),
\label{eq:rateAnalytic}
\end{equation}
where
$\epsilon_{\rm BW} \equiv GM_\bullet/ (2r_{\rm BW})$.  Using the analytic diffusion coefficients from Appendix \ref{app:LC}, we find that in the ultrasteep regime $(\gamma > 9/4)$,
\be \dot{N}_\rho(t) \propto t^{-(4\gamma-9)/(2\gamma-3)}/\ln t. \label{eq:powerlaw} \ee
A potentially important complication to any scenario involving extremely dense stellar systems is the role of direct physical collisions between stars \citep[e.g.][]{FreitagBenz02}.  In some portions of parameter space, direct collisions could erode an overdense cusp on timescales shorter than the local relaxation time.  For cusps consisting of a single species of stars, the per-star collision rate is
\begin{equation}
\dot{N}_{\rm coll}(r)= \pi\sqrt{3} R_\star^2 \frac{\rho (r)}{M_\star} \sigma(r) \left(1 + \frac{4GM_\star}{3R_\star \sigma^2(r)} \right)
\end{equation}
Comparing the relaxation time to the characteristic stellar collision time $t_{\rm coll}=\dot{N}_{\rm coll}^{-1}$, we find
\begin{equation}
\frac{t_{\rm r}}{t_{\rm coll}}=\pi \sqrt{3}(1+\gamma)^{3/2}k(\gamma) \frac{\sigma^4(r) R_\star^2}{G^2 \langle m_\star^2 \rangle \ln \Lambda} \left(1 + \frac{4GM_\star}{3R_\star \sigma^2(r)} \right). \label{eq:tColl}
\end{equation}
When $t_{\rm r}/t_{\rm coll} \gtrsim 1$, collisional erosion of the stellar cusp dominates relaxational evolution and standard LC theory will strongly overestimate TDE rates.  In general, however, $t_{\rm r}/t_{\rm coll} \gtrsim 1$ only on very small scales, of $r\lesssim 3\times 10^{-3}~{\rm pc}$ ($r \lesssim 3\times 10^{-2}~{\rm pc}$) for $M_\bullet = 10^6 M_\odot$ ($M_\bullet = 10^7 M_\odot$).  As we shall see in \S \ref{sec:overdenseNumerical}, these scales are only relevant for extremely early periods of post-starburst evolution.

\begin{figure}
  \includegraphics[width=85mm]{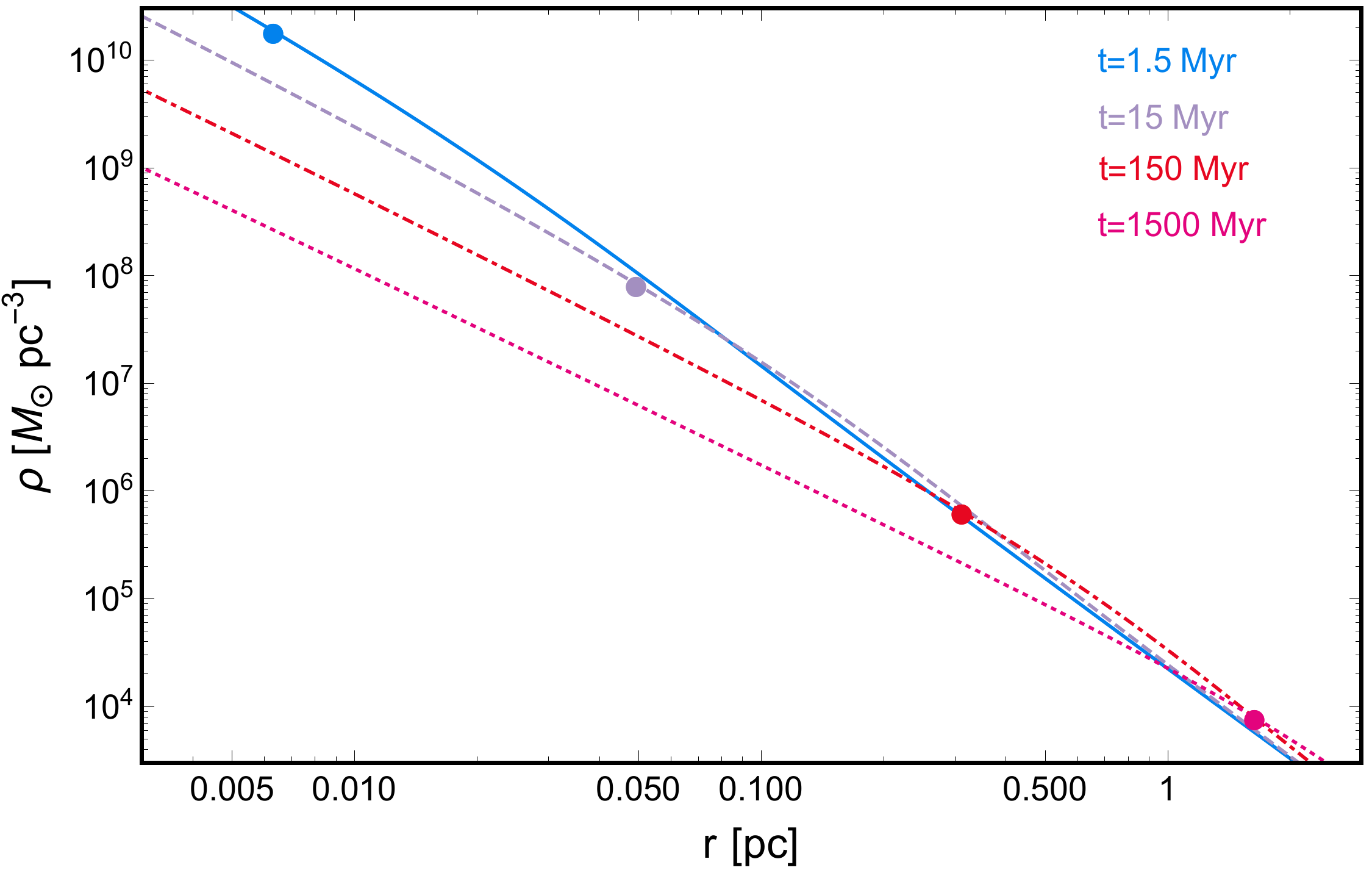}
  \caption{\label{fig:rhoOverdense} Time evolution of an initially ultrasteep $\rho \propto r^{-2.75}$ stellar cusp using the \textsc{PhaseFlow} code.  The blue solid, purple dashed, red dot-dashed, and pink dotted curves correspond to post-starburst ages $t=1.5~{\rm Myr}$, $t=15~{\rm Myr}$, $t=150~{\rm Myr}$, and $t=1500~{\rm Myr}$, respectively. Here we consider a cusp of equal-mass ($M_\star = 1M_\odot$) stars around a $M_\bullet=10^6 M_\odot$ SMBH.  We mark the Bahcall-Wolf radius $r_{\rm BW}$ with a square on each snapshot in the top panel.}
\end{figure}

\begin{figure}
  \includegraphics[width=85mm]{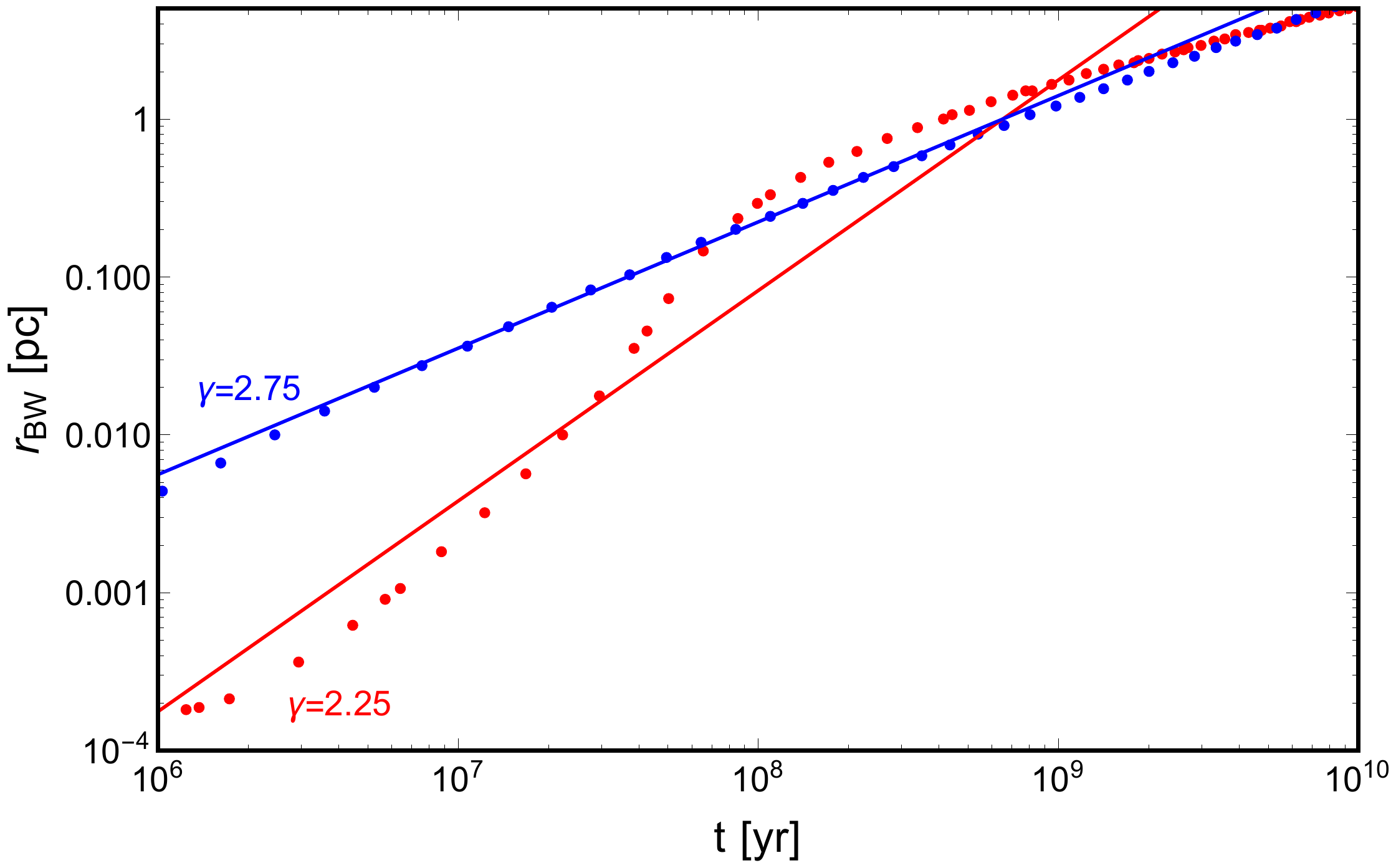}
  \caption{\label{fig:rBW} The ``Bahcall-Wolf'' break radius $r_{\rm BW}$ as a function of time, comparing our numerical calculations (dots) to the analytic approximation in Eq.~\ref{eq:rBW} (solid lines; here we have numerically calibrated the $k(\gamma)$ normalization factor).  The time evolution of the numerical Fokker-Planck solutions is well-approximated by our analytic estimate, though the normalization differs by a factor of a few.  For the numerical points, this radius is defined as the location where the local logarithmic density slope equals the average of the BW profile's $\gamma=7/4$ and the initial power-law slope.  Eq. \ref{eq:rBW} does not perform as well at estimating the time evolution of the marginally ultrasteep $\gamma=2.25$ initial conditions.}
\end{figure}

\subsection{Numerical Evolution of Stellar Overdensities}
\label{sec:overdenseNumerical}
To validate the above analytic prescription, we solve the time
dependent, isotropic Fokker-Planck equation in energy space. This can be written
in flux-conserving form as
\begin{equation}
\frac{\partial f(\epsilon, t)}{\partial t} = -\frac{\partial}{\partial
  \epsilon} \left[ D_{\epsilon \epsilon} \frac{\partial f(\epsilon,
    t)}{\partial \epsilon} + D_{\epsilon} f(\epsilon, t) \right]
-\frac{f(\epsilon, t)}{t_{\rm LC} (\epsilon, t)},
\label{eq:fpen}
\end{equation}
where $D_{\epsilon}$ and $D_{\epsilon \epsilon}$ are the first and
second order diffusion coefficients respectively (see
e.g. \citealt{Merritt13}).
The last term approximates the loss of stars due to angular momentum diffusion into the
loss cone for each energy bin (see eq. 13 in \citealt{Vasiliev17}). For a
single component system the diffusion coefficients can be written as
integrals over the distribution function:
\begin{align}
  D_{\epsilon\epsilon}=& 16 \pi^2 G^2 m \ln \Lambda \left[ h(\epsilon) \int_0^{\epsilon}
    f(\epsilon') d\epsilon' + \int_\epsilon^\infty f(\epsilon') h(\epsilon')\right] \\
  D_{\epsilon}=& -16 \pi^2 G^2 m \ln \Lambda  \int_{\epsilon}^{\infty}
  f(\epsilon') g(\epsilon') d\epsilon',
\end{align}
where $h(\epsilon)$ is the phase volume and $g(\epsilon)={\rm d}h(\epsilon)/{\rm d}\epsilon$ is the density of states.

To solve this equation we use the \textsc{PhaseFlow} code
(Vasiliev 2017).\footnote{\textsc{PhaseFlow} is part of the \textsc{Agama} library for galaxy modelling, available from
https://github.com/GalacticDynamics-Oxford/Agama/.  Note that \textsc{PhaseFlow} uses the phase
  volume $h(\epsilon)$ rather than energy as the independent variable.}  Our initial
conditions, motivated by a rapid burst of star formation, are a power
law stellar density profile of 1 $\Msun$ stars with a range of initial power law slopes $\gamma$ and influence radii $r_{\rm infl, 0}$. To avoid divergences in the total mass and
stellar potential we introduce an exponential cut-off in the density
at 1000 pc and a small $r^{-1}$ core inside of $10^{-4}$ pc. The inner
boundary is absorbing, with the distribution function having a fixed
value of zero. The outer boundary condition corresponds to zero flux.

Fig.~\ref{fig:rhoOverdense} shows the relaxation of an ultrasteep
$r^{-2.75}$ stellar cusp from \textsc{PhaseFlow}. As expected, the stellar density
relaxes to an $r^{-7/4}$ Bahcall-Wolf cusp from the inside-out.
In Fig. \ref{fig:rBW}, we plot the evolution of the Bahcall-Wolf transition radius $r_{\rm BW}$ versus time.  Eq. \ref{eq:rBW} captures the time-dependent evolution of the
Bahcall-Wolf radius for $r^{-2.25}$ and $r^{-2.75}$ profiles from
$\sim 10^6$ years until $r_{\rm BW}$ approaches the influence radius (because $r_{\rm BW} \propto t^{1/(\gamma-3/2)}$, the $r^{-2.25}$ profile's break radius deviates from analytic predictions much sooner than does the $r^{-2.75}$ profile's). The agreement between
analytic theory and numerical results for the power-law slope of
$r_{\rm BW}(t)$ is quite good in the ultrasteep regime, though the normalization of $r_{\rm
  BW}$ would differ by a factor of a few between these approaches if we computed $k(\gamma)$ using the local diffusion coefficient of Appendix \ref{app:LC}.  The
agreement is worse for shallower profiles. However, the comparison is more
ambiguous in this case, as the inner and outer
density profiles are similar.

\begin{figure}
\includegraphics[width=85mm]{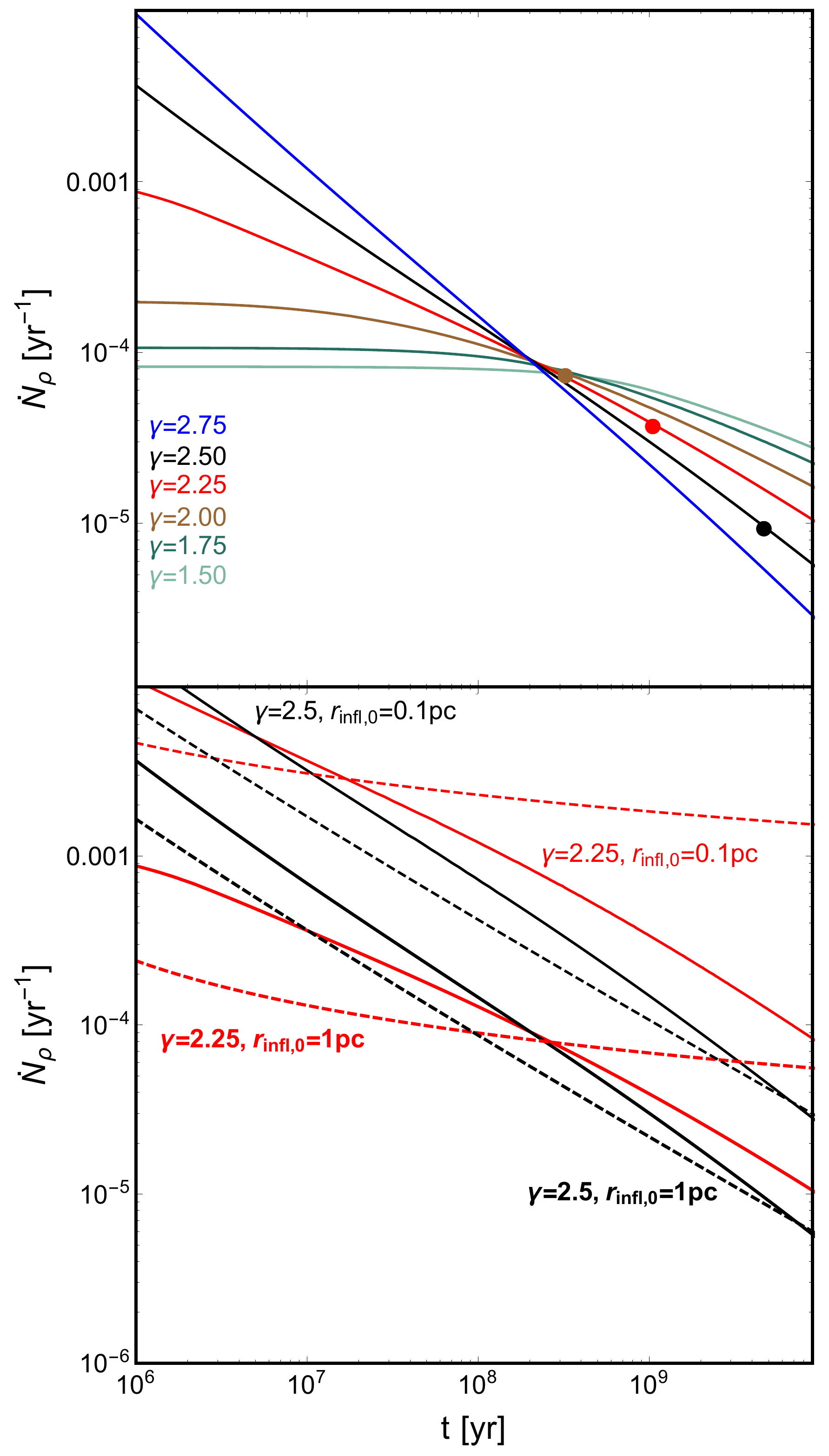}
\caption{\label{fig:agamaDTD} TDE rates as functions of time, for
  different initially overdense stellar density profiles, as calculated
  numerically by \textsc{PhaseFlow} around a $M_\bullet=10^6 M_\odot$
  SMBH. {\it Top panel:} the power law slope of the density profile is
  varied, while the initial influence radius is fixed at 1 pc. Circles indicate the time when the initial influence radius $r_{\rm infl,0}$ is overtaken by the expanding $r_{\rm BW}$, for initial $\gamma>7/4$.  {\it Bottom
    panel:} the effect of decreasing the influence radius to 0.1 pc
  is shown for ultrasteep profiles.  Black curves are ultrasteep $\gamma=2.5$ profiles, while red curves are more marginal $\gamma=2.25$ profiles.  Thick (thin) lines show $r_{\rm infl,0}=1~{\rm pc}$ ($r_{\rm infl,0}=0.1~{\rm pc}$).  Numerical solutions are solid lines, while the analytic
  rates from Eq.~\ref{eq:rateAnalytic} are plotted as
  dashed lines.  As in Fig. \ref{fig:rBW}, a good fit in time evolution is seen for ultrasteep profiles, but this breaks down for marginally ultrasteep $\gamma=2.25$ cusps.  The analytic normalization is off by a factor of a few from numerical results.}
\end{figure}

The top panel of Fig.~\ref{fig:agamaDTD} shows the TDE rate as a
function of time for different initial density slopes $\gamma$.  The bottom
panel shows how the TDE rate depends on the initial influence radius
and compares TDE rates extracted from \textsc{PhaseFlow} with our analytic approximation (Eq.~\ref{eq:rateAnalytic}). The analytic prescription
reproduces the time evolution of the TDE rate well for the
$r^{-2.5}$ and $r^{-2.75}$ profiles. However, the TDE rate from
Eq.~\ref{eq:rateAnalytic} has a slightly shallower slope at late times
than do the numerical results, and the normalization is off by a factor of a few, due to the crudeness of the assumption that $\dot{N}_\rho \sim \epsilon_{\rm BW}\mathcal{F}(\epsilon_{\rm BW})$. Eventually, the Bahcall-Wolf radius
approaches the initial influence radius $r_{\rm infl,0}$ (the dots on each curve in the top
panel of Fig.~\ref{fig:agamaDTD} mark the time when $r_{\rm BW}=r_{\rm infl,0}$). Afterwards, the
evolution of the TDE rate is driven by the expansion of the star
cluster due to the consumption of stars, an effect not captured by the
analytic model. The analytic prescription is also generally inaccurate for
marginally ultrasteep initial density profiles, i.e. $\gamma\approx 2.25$, because for these marginal $\gamma$ values the LC flux is less sharply peaked at $\epsilon_{\rm BW}$, and the Bahcall-Wolf radius more swiftly overtakes the initial influence radius.

\begin{figure}
\includegraphics[width=8.5cm]{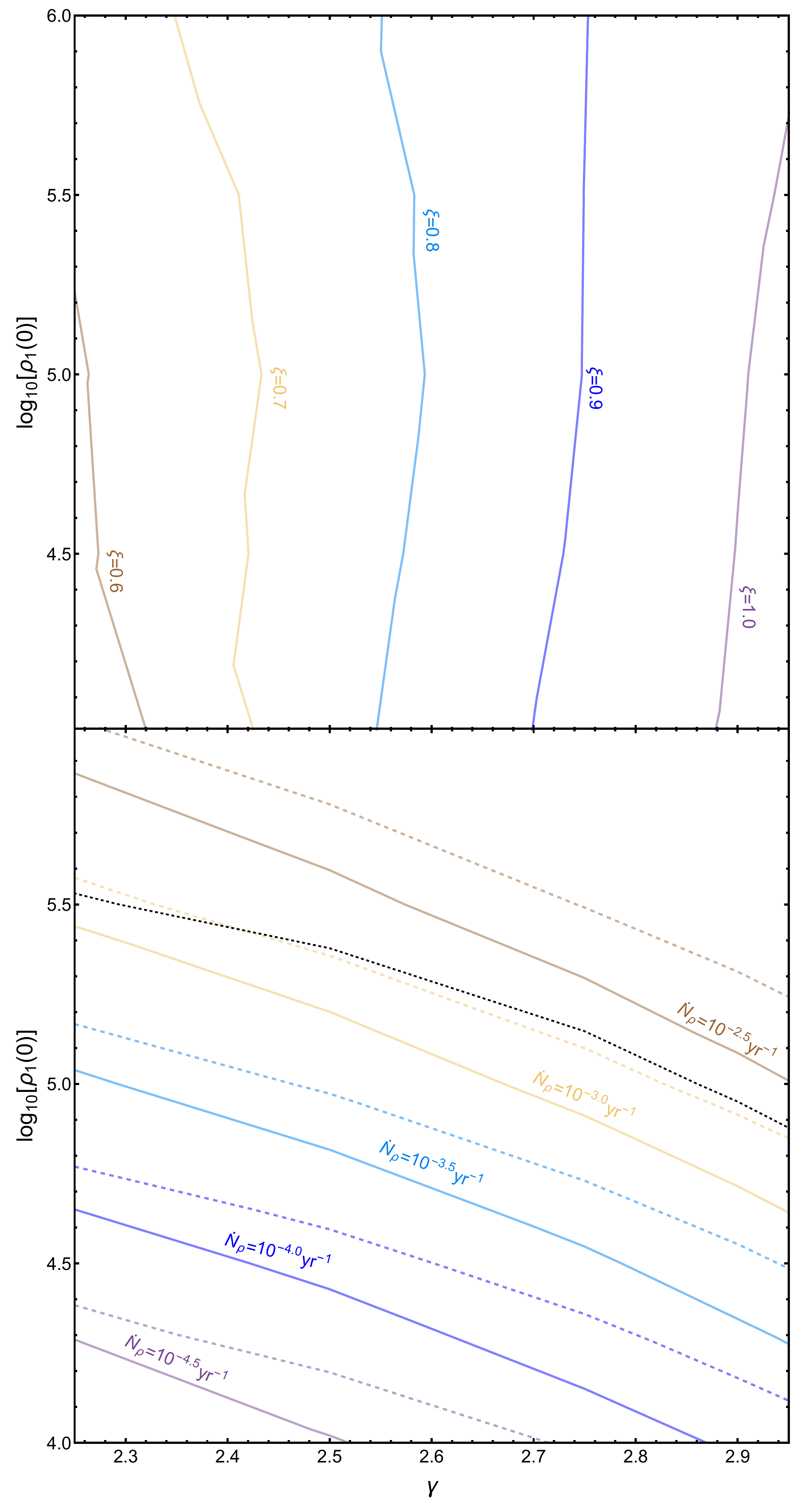}
\caption{\label{fig:rateContours} \emph{Top panel:}  The time evolution of the TDE rate is well approximated by a power-law $\dot{N}_{\rho} \propto t^{-\xi}$, for initial stellar density profiles $\rho \propto r^{-\gamma}$ in the ultra-steep regime $\gamma \ge 9/4$ (eq.~\ref{eq:powerlaw}).  Here we show contours of $\xi$ as a function of $\gamma$, measured by averaging our numerical results over the time interval $10^8~{\rm yr}<t<10^{10}$ yr. \emph{Bottom
    panel:} Contours of the TDE rate at $t= 100$ Myr (\emph{solid lines}) and
  200 Myr (\emph{dashed lines}) as a function of the initial stellar density profile $\gamma$ and the normalization of the initial stellar density profile, $\rho_1(0) \equiv \rho(t=0, r = 1$ pc$)$.  Above the black dotted line, the central SMBH mass would double over $10$ Gyr if it accreted half of each disrupted star (the effects of which are not included in our calculations). }
\end{figure}

We better quantify the time evolution of the TDE rate by calculating the average power law index for our Fokker-Planck results (i.e. $\dot{N}_\rho \propto t^{-\xi}$, where $\xi$ is averaged between $10^8~{\rm yr}$ and $10^{10}~{\rm yr}$).  The top panel of Fig.~\ref{fig:rateContours} shows contours of constant $\xi$ for different values of the initial slope $\gamma$ of the stellar density profile, and its initial normalization at 1 parsec, $\rho_1(0)\equiv\rho(t=0,r = 1$ pc$)$.  The bottom panel of the same figure shows contours of TDE rate at $t = 100$ Myr in the same parameter space.

Taken together, Figs. \ref{fig:agamaDTD} and \ref{fig:rateContours} disentangle the loss cone behavior of ultrasteep and overconcentrated cusps.  The top panel of Fig. \ref{fig:agamaDTD} shows strikingly similar rates at post-starburst ages $t \sim 10^8~{\rm yr}$, implying that - given a constant value of $r_{\rm infl,0}$ - ultrasteep profiles alone are not sufficient to explain the PSP in typical galaxies.  In contrast, the bottom panel of Fig. \ref{fig:rateContours} shows that $\dot{N}_\rho(10^8~{\rm yr})$ has a more nontrivial dependence on $\gamma$ if it is instead $\rho_1(0)$ that is held constant, and the bottom panel of Fig. \ref{fig:agamaDTD} shows how varying $r_{\rm infl,0}$ into the overconcentrated regime brings post-starburst rates up into agreement with observations.  The difference between these two figures highlights how $r_{\rm infl,0}$ becomes a poor metric of overconcentration in the ultrasteep regime.  Large changes in initial $r_{\rm infl}$ produce only modest variation in $\rho_1(0)$, and it is the latter that is more predictive of the TDE rate during post-starburst ages $t\gtrsim 10^8~{\rm yr}$.  In most overconcentrated nuclei, the initial influence radius $r_{\rm infl,0}$ expands significantly during $10^{10}~{\rm yr}$ of relaxational evolution.

In summary, overconcentrated nuclei with low $r_{\rm infl,0}$, or high $\rho_1(0)$, are able to reproduce observed magnitudes of TDE rates in E+A galaxies.  Several of the curves in the bottom panels of Figs. \ref{fig:agamaDTD} and \ref{fig:rateContours} are capable of reproducing the inferred TDE rates in E+A galaxies \citep[$\dot{N}\sim 10^{-3}~{\rm yr}^{-1}$;][]{French+16}.  In general, increasing the density slope $\gamma$ increases the initial TDE rate but also increases the steepness of the DTD, while diminishing $r_{\rm infl}$ (or increasing $\rho_1(0)$) increases the early-time TDE rate.  The overall slope $\xi$ of the delay time distribution $\dot{N}_\rho(t)$ is controlled by the initial power law slope $\gamma$.  When $\gamma$ is in the ultrasteep regime, our analytic formula for the DTD (Eq. \ref{eq:powerlaw}) describes its time evolution well.  We defer a more detailed discussion of these considerations for \S \ref{sec:DTD}.

\section{Radial Velocity Anisotropies}
\label{sec:aniso}

An alternative dynamical explanation for the PSP is a radial orbit
bias among the stars formed in the center of a galaxy during a
starburst\footnote{The effect of radial
  anisotropies on TDE rates was first commented on in
  \citet{MagorrianTremaine99}.}.  Because stars on nearly radial
orbits require far less time to diffuse through angular momentum space
into the SMBH loss cone, a stellar cluster born with a radial velocity
anisotropy will see an initially elevated TDE rate (relative to an
otherwise equivalent cluster with an isotropic velocity field).
However, the same two-body relaxation process that feeds stars to the SMBH also acts to isotropize the velocity distribution.
Tangential velocity anisotropies have been explored previously as
a mechanism to suppress TDE rates \citep{LezhninVasiliev15,
  LezhninVasiliev16}; these investigations generally find that their
anisotropies wash out in a fraction of an energy relaxation time.  We expect that an initially radial velocity-space bias will also isotropize in a fraction of a relaxation time.  

It is not obvious why a starburst would bias nuclear orbits in a radial sense (indeed, the disks of young stars seen in the Galactic Center have a pronounced tangential anisotropy, e.g. \citealt{Bartko+09}).  We speculate that the infall and tidal disruption of young massive clusters could inject a fresh population of stars onto radial orbits in a post-starburst galactic nucleus, but modeling the full history of nuclear starbursts is beyond the scope of this paper.  In this section, we simply investigate the self-consistent evolution of DFs with an initial radial bias, calculating the magnitude and duration of any enhancement to TDE rates therein.

We model the time evolution of the TDE rate in a galaxy with anisotropic stellar velocities by solving the time-dependent, one-dimensional Fokker-Planck equation in angular momentum space.  By assumption, stars are fixed in bins of orbital energy, but allowed to diffuse through angular momentum space in a random walk.  This process is captured by the orbit-averaged Fokker-Planck equation \citep{MerrittWang05},
\begin{equation}
\frac{\partial f}{\partial \tau} = \frac{1}{4 j} \frac{\partial}{\partial j}\left(j \frac{\partial f}{\partial j} \right), \label{eq:FPJ}
\end{equation}
where $j \equiv J/J_{\rm c}(\epsilon)$ is the dimensionless angular momentum (here $J_{\rm c}(\epsilon)$ is the specific angular momentum of a circular orbit), $\tau \equiv \bar{\mu}(\epsilon) t \sim t / t_{\rm r}$ is a dimensionless time, and $\bar{\mu}(\epsilon)$ is the orbit-averaged angular momentum diffusion coefficient (see \S \ref{sec:overdense}).  We again assume a spherical geometry for the star cluster, and a separable distribution function $f(\epsilon, j) = f_\epsilon(\epsilon) f_{\rm j}(j)$.  The latter assumption - evolving $j$ at fixed $\epsilon$ - is valid on timescales short compared to the energy relaxation time because angular momentum relaxation is much faster than energy relaxation for the radial orbits that concern us.  However, Eq. \ref{eq:FPJ} will break down at late times, once either
\begin{enumerate}
\item An energy relaxation time has passed, and $f_\epsilon(\epsilon)$ can no longer be treated as static, or
\item An order unity fraction of stars with specific energy $\epsilon$ have drained into the SMBH loss cone.
\end{enumerate} 
The outer boundary condition for Eq. \ref{eq:FPJ} is of the Neumann type to prohibit flux of stars through $j=1$:
\begin{equation}
\frac{\partial f_{\rm j}}{\partial j}\bigg\rvert_{j=1} = 0.
\end{equation}
The inner boundary condition depends greatly on the dimensionless diffusivity parameter $q(\epsilon) = P(\epsilon)\bar{\mu}(\epsilon)/j^2_{\rm LC}(\epsilon) \approx \Delta J^2/J_{\rm LC}^2$, where $j_{\rm LC} \equiv \sqrt{2GM_\bullet R_{\rm t}}/J_{\rm c}(\epsilon)$ and $P(\epsilon)$ is the orbital period.  In the strongly diffusive regime ($q\ll 1$, also known as the ``empty loss cone'' regime), stars are immediately destroyed once reaching $j=j_{\rm LC}$ and the inner boundary condition is of the Dirichlet type: $f(j_{\rm LC}, t) = 0$.  However, in the pinhole regime ($q\gg1$, also known as the ``full loss cone'' regime), relaxation can cause stars to wander to values of $j<j_{\rm LC}$ many times per orbit and the loss cone is not empty.  In this case the inner boundary condition is of the Robin type \citep{VasilievMerritt13, LezhninVasiliev15},
\begin{equation}
f_{\rm j}(j_{\rm LC}) = \frac{\alpha(q) j_{\rm LC}}{2} \frac{\partial f_{\rm j}}{\partial j}\bigg\rvert_{j=j_{\rm LC}},
\end{equation}
where $\alpha(q) \equiv (q^2+q^4)^{1/4}$.

Following standard conventions, we parametrize the velocity anisotropy as
\begin{equation}
\beta \equiv 1-\frac{T_\perp}{2T_\parallel}, \label{eq:beta}
\end{equation}
where $T_\perp$ and $T_\parallel$ are the kinetic energies of tangential and radial motion, respectively, integrated across the stellar distribution function.  If all orbits are purely radial, then $\beta = 1$; if all orbits are purely tangential, then $\beta = -\infty$.  An isotropic distribution corresponds to $\beta=0$, so for the remainder of this section we focus on scenarios where $\beta>0$, corresponding to a radial velocity bias.  This parametrization of anisotropy maps onto the distribution function simply, with $f_j(j) \propto j^{-2\beta}$.  As initial conditions, we therefore take\footnote{In order for our initial conditions to match the outer boundary condition, we select initial conditions that deviate very slightly from Eq. \ref{eq:ICs}.  Specifically, for $j>0.999$, we set the initial DF to a constant value equal to $f(0.999, t=0; \epsilon)$.}
\begin{equation}
f_{\rm j}(j, t=0) = \frac{1-\beta_0}{1-j_{\rm LC}^{2-2\beta_0}}j^{-2\beta_0}, \,\,\,j_{\rm LC} < j \le 1\label{eq:ICs}.
\end{equation}
For convenience we have normalized the distribution function to unity, though this choice is inessential, as our interest is in studying the {\it relative} enhancement of the TDE rate (and its time evolution) due to a radial velocity bias in comparison to an isotropic velocity distribution.

For simplicity, we also assume in this section that the initial anisotropy $\beta_0$ is constant across all orbital energies $\epsilon$.  An upper limit on the allowed range of $\beta_0$ is set by the radial orbit instability (ROI).  If the radial anisotropy exceeds some critical value $\beta_{\rm ROI}$, the stellar system will become unstable to nonspherical perturbations, and the geometry of the star cluster will evolve into a nonspherical configuration \citep{Antonov73, Henon73}.  Such a configuration will enhance TDE rates further (at least temporarily) through collisionless effects \citep{MerrittPoon04}, but because the orbital dynamics and evolution of the distribution will become considerably more complicated, we defer an investigation of this scenario to future work.  Although analytic criteria have suggested that $\beta_{\rm ROI}\approx 0.43$ \citep{PolyachenkoShukhman81}, numerical orbit integrations find stability up to larger anisotropies, with $\beta_{\rm ROI} \approx 0.6$ \citep{MerrittAguilar85}.  While the precise ROI threshold depends somewhat on the shape of the potential (and the distribution of anisotropies, which can of course differ from the power law ansatz in Eq. \ref{eq:beta}), we explore a large range of $\beta_0$, with the caution that the most anisotropic models may not be self-consistent.  Other dynamical instabilities may arise for non-monotonic $f_j(j)$ \citep{Polyachenko+08}, but we defer a discussion of this constraint for future work.

Since the approximate timescale for angular momentum relaxation to occur is $t_{\rm j}(j) \approx j^2 t_{\rm r}$, we expect radially biased initial conditions to relax from the inside out in angular momentum space, analogous to energy space relaxation in ultrasteep cusps.  We can likewise compute the range of angular momenta that have collisionally relaxed at some time $t$ after the initialization of radially anisotropic initial conditions.  In analogy to the ``Bahcall-Wolf energy'' $\epsilon_{\rm BW}$ of \S \ref{sec:overdense}, we find a ``Cohn-Kulsrud angular momentum''
\begin{equation}
j_{\rm CK} \approx \sqrt{\tau
}, \label{eq:jCK}
\end{equation}
interior to which the angular momentum DF should reach a steady state solution similar to \citealt{CohnKulsrud78} (here $\tau$ is the dimensionless time introduced after Eq.~\ref{eq:FPJ}).

\begin{figure}
\includegraphics[width=85mm]{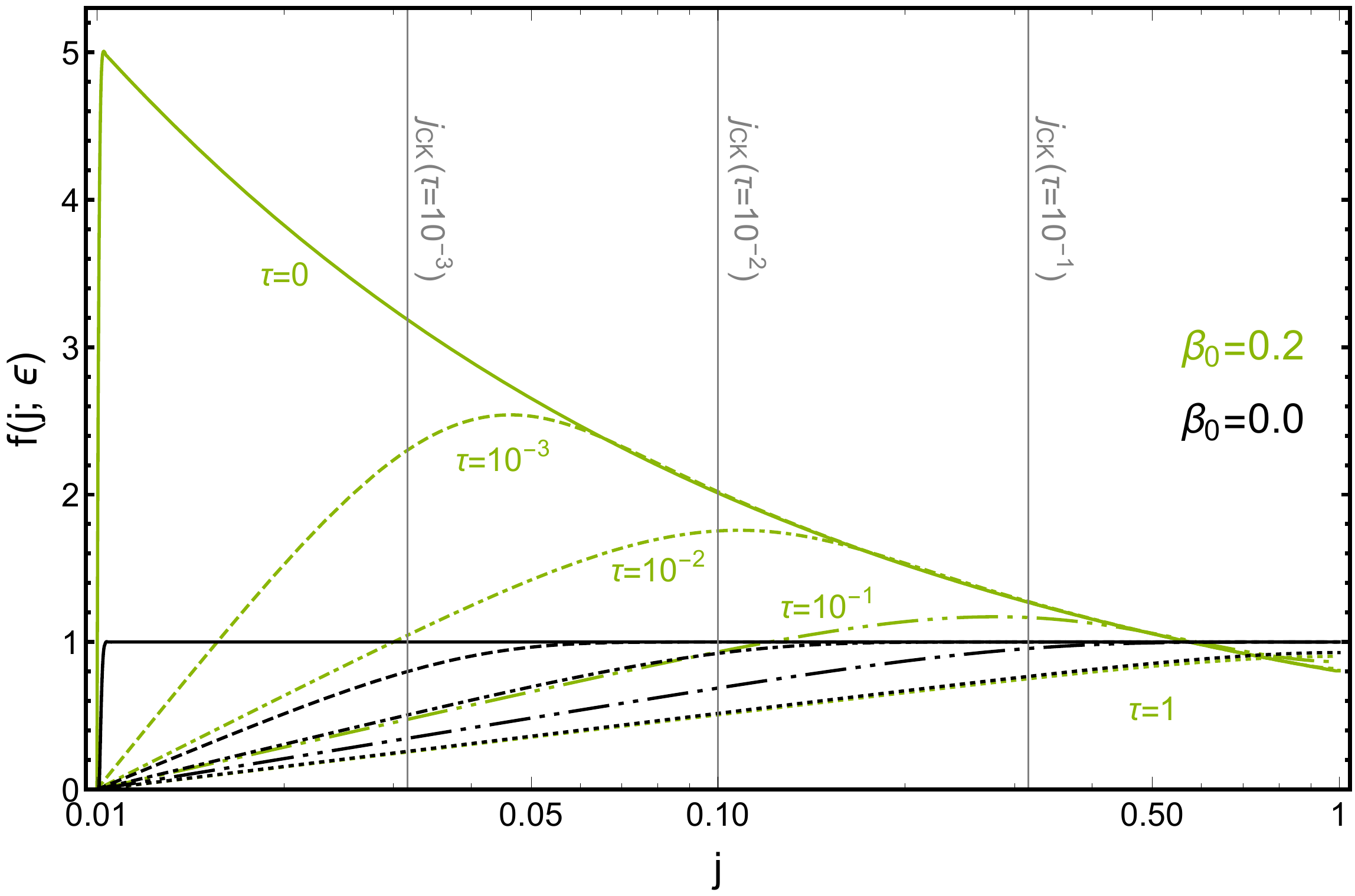}
\caption{Evolution of the distribution function $f(j;\epsilon)$ as a function of the dimensionless angular momentum $j$ at fixed energy $\epsilon$, shown for different snapshots in dimensionless time $\tau = 0$ (solid), $\tau = 10^{-3}$ (dashed), $\tau = 10^{-2}$ (dot-dashed), $\tau=10^{-1}$ (dot-dot-dashed) and $\tau = 1$ (dotted).  Black lines correspond to a isotropic stellar cluster ($\beta_0 = 0$), while light lines show a case with moderate initial radial anisotropy ($\beta_0 = 0.2$).  All cases assume a loss cone boundary condition $j_{\rm LC} = 10^{-2}$ and a diffusive regime of LC repopulation ($q=0.01$).  The last snapshot, though technically beyond the domain of validity of our one-dimensional Fokker-Planck calculation, is shown for completeness.}
\label{fig:anisoEvol}
\end{figure}

Fig. \ref{fig:anisoEvol} shows the time evolution of the distribution function for an initially radially-biased profile with $\beta_0=0.2$, as compared to the standard isotropic case $\beta_0 = 0$, for $j_{\rm LC}=10^{-2}$ (an unusually large value chosen only for convenience of plotting).  We find that the radially biased-initial conditions result in more stars being retained on low angular momentum orbits until roughly a relaxation time has passed (after which point the one-dimensional Fokker-Planck approach used is no longer valid).  The time evolution of $f(j, t;\epsilon)$ is to zeroth order captured by Eq. \ref{eq:jCK}; radially anisotropic initial conditions approach their logarithmic, steady state solution \citep{CohnKulsrud78} interior to a value of $j$ that is within a factor of two of $j_{\rm CK}$.  For $j\gtrsim j_{\rm CK}$, the DF has not greatly evolved from its initial conditions.  We therefore expect an enhanced flux of stars into the loss cones for all times $\tau < 1$ during which our calculation is valid.

\begin{figure}
\includegraphics[width=85mm]{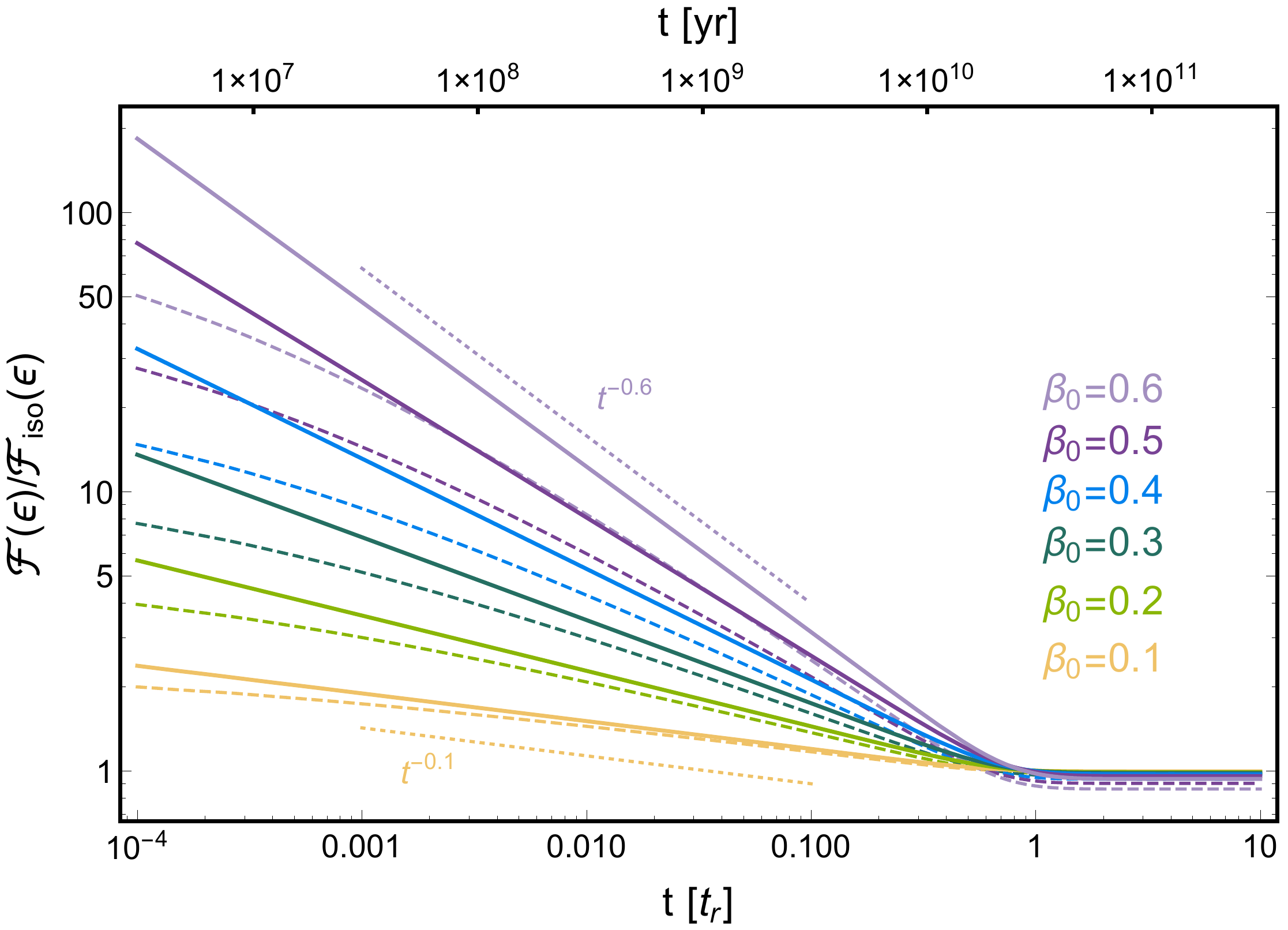}
\caption{Time evolution of the flux of stars into the SMBH loss cone, $\mathcal{F}$, normalized to the equivalent stellar flux for isotropic initial conditions, $\mathcal{F}_{\rm iso}$.  Curves represent numerical solutions to Eq. \ref{eq:FPJ}.  Time on the lower horizontal axis is normalized to the relaxation time $t_{\rm r}$; on the upper axis we show the physical passage of time for a low-mass cusp galaxy with $M_\bullet=10^{6.5} M_\odot$ and $\gamma=1.6$.  We consider a range of different anisotropic initial conditions, with $\beta_0=0.1$ (yellow), $0.2$ (light green), $0.3$ (dark green),  $0.4$ (blue), and $0.5$ (purple), and $0.6$ (lavender).  The largest $\beta_0$ values may trigger the onset of the radial orbit instability, limiting the validity of those calculations.  Solid lines represent energy bins near the influence radius from where most TDEs are sourced ($j_{\rm LC} = 10^{-5}$) while dashed lines represent much more tightly bound stellar orbits ($j_{\rm LC} = 10^{-2}$); in both cases we consider strongly diffusive ($q=0.01$) LC repopulation.  Dotted lines show analytic power law estimates $\dot{N} \propto t^{-\beta_0}$ (Eq. \ref{eq:betaLaw}), which are generally in excellent agreement with our numerical solutions.  }
\label{fig:anisoFluxNormalized}
\end{figure}

The instantaneous loss cone flux is calculated as 
\begin{equation}
\mathcal{F}(t; \epsilon) = 2 \pi^2 \bar{\mu}(\epsilon) P(\epsilon) J_{\rm c}^2(\epsilon) f_{\epsilon}(\epsilon) \left(j \frac{\partial f_{\rm j}(j, t)}{\partial j} \right)_{j = j_{\rm LC}}. \label{eq:LCFlux}
\end{equation}
For our initial conditions, the TDE rate from a given energy bin, $\mathcal{F}(t; \epsilon)$, will be proportional to $f_{\rm j}(j_{\rm CK}, t)$ (as it is the Cohn-Kulsrud angular momentum which determines the slope of the DF into the loss cone), implying the surprisingly simple time evolution 
\begin{equation}
\dot{N}_\beta(t) \propto t^{-\beta_0}. \label{eq:betaLaw}
\end{equation}
Fig. \ref{fig:anisoFluxNormalized} shows the time evolution of the diffusive loss cone flux enhancement $\mathcal{F}(t; \epsilon)/\mathcal{F}_{\rm iso}(t;\epsilon)$ for different degrees of anisotropy $\beta_0 = 0.1-0.6$ and two assumptions about the loss-cone angular momentum ($j_{\rm LC} = 10^{-5},10^{-2}$), normalized to the otherwise equivalent flux for isotropic ($\beta_0 = 0$) initial conditions, $\mathcal{F}_{\rm iso}(t;\epsilon)$.  Larger values of $j_{\rm LC}$ correspond to stars tightly bound to the SMBH (high values of $\epsilon$), while smaller values represent those near the radius of influence of a low-mass SMBH (lower values of $\epsilon$).  Results for the pinhole regime are very similar and are not shown.  Note that all energy dependent quantities in $\mathcal{F}(t; \epsilon)/\mathcal{F}_{\rm iso}(t;\epsilon)$ cancel out (aside from the value of $j_{\rm LC}$ used to solve Eq. \ref{eq:FPJ}).


In Fig. \ref{fig:anisoFluxNormalized}, we see that an initial radial anisotropy enhances the TDE rate by a factor $\approx 2-200$ compared to the otherwise equivalent isotropic case.  The rate declines as a power law in time that (for $j_{\rm LC}\ll 1$) is well described by Eq. \ref{eq:betaLaw}.  The upper horizontal axis of Fig. \ref{fig:anisoFluxNormalized} shows time in years for a $10^{6.5} M_\odot$ SMBH with ``typical'' values of $r_{\rm infl}=1.5~{\rm pc}$ and $\gamma = 0.61$ (these values are estimated by fitting power laws to the distributions of $\gamma$ and $r_{\rm infl}$ estimated from the galaxy sample of \citealt{StoneMetzger16}; see \S \ref{sec:DTD}).  For a very high initial anisotropy, $\beta_0 = 0.6$, we find that a system of age $t=10^8~{\rm yr}$ will have a rate enhancement $\approx 50$.  If the ROI limits $\beta_0$ to values $\le 0.5$, the enhancement is reduced to $\le 20$.

The TDE rate under radially-biased conditions eventually falls below the isotropic one as stars are depleted from the DF; this occurs after a time $\tau \sim 1$.  However, this also marks the point after which the one-dimensional Fokker-Planck equation can no longer be trusted because energy space diffusion  has become important.  By this point, the distribution is largely isotropic and further evolution in the TDE rate occurs on the longer timescale for energy relaxation.  

So far we have considered only the evolution of the angular momentum distribution $f_{\rm j}(j)$ at fixed orbital energy $\epsilon$.  In typical, non-ultrasteep galaxies ($\gamma<9/4$) with isotropic DFs, the integrated LC flux is sharply peaked at energies $\epsilon \approx \epsilon_{\rm crit}$ \citep{WangMerritt04}; naively, this might imply that we could estimate the TDE rate in a radially anisotropic galaxy by considering the evolution of $f(j; \epsilon_{\rm crit})$ alone.  To test this hypothesis we integrate Eq. \ref{eq:FPJ} numerically in many different bins of orbital energy across a distribution function $f(\epsilon)$.  The total TDE rate $\dot{N}_\beta(t) = \int \mathcal{F}(t; \epsilon){\rm d}\epsilon$, where $\mathcal{F}(t; \epsilon)$ is given by Eq. \ref{eq:LCFlux}.  

\begin{figure}
\includegraphics[width=85mm]{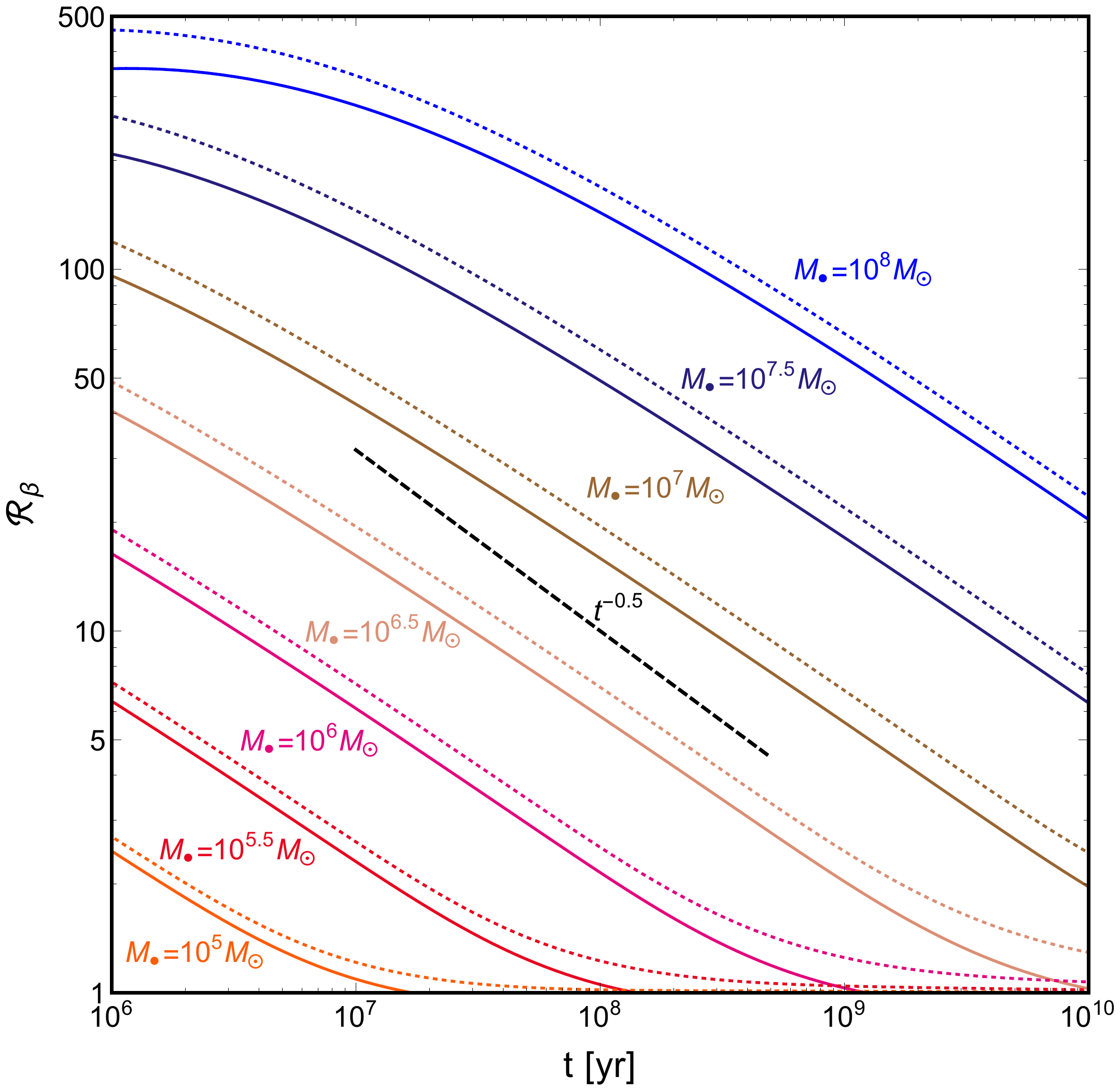}
\caption{TDE rate enhancements $\mathcal{R}_\beta(t)\equiv \dot{N}_\beta(t) / \dot{N}_{\rm iso}(t)$ in idealized galactic models with initial radial bias $\beta_0 = 0.5$, plotted against physical post-starburst age $t$.  Solid lines compute TDE rates $\dot{N}_\beta(t)$ by integrating across a DF $f_\epsilon(\epsilon)$ computed from Eddington's formula, while dotted lines use a DF $f_\epsilon(\epsilon)$ that is derived self-consistently.  The difference is small, $\approx 20\%$.  Different colors correspond to different SMBH and host galaxy masses; the SMBH masses $M_\bullet$ are labeled in the figure.  The dashed black line is a power law going as $t^{-0.5}$, as in Eq. \ref{eq:betaLaw}.  This analytic estimate for the anisotropic DTD is still a reasonable approximation, but it is less precise here than it was in Fig. \ref{fig:anisoFluxNormalized}, where only single bins of orbital energy $\epsilon$ were considered.}
\label{fig:anisoGalaxies}
\end{figure}

We construct a grid of galaxy models varying $M_\bullet$ between $10^5 M_\odot$ and $10^8 M_\odot$, and initial anisotropy parameter $\beta_0$ between $0.0$ and $0.6$.  Each stellar system is idealized as a Dehnen model \citep{Dehnen93, Tremaine+94}: a spherical potential-density pair with a density profile $\rho(r)$ that is a smoothly broken power law.  The inner region controls the TDE rate and has a power law density slope $\gamma$ determined by least-squares fitting to the galaxy sample of \citet{StoneMetzger16}.  The fit we find is $\gamma = 2.4997 - 0.1371\log_{10}(M_\bullet / M_\odot)$; likewise, we set the break radius by assuming the galaxy's total stellar mass $\mathcal{M}_\bullet = 10^3 M_\bullet$, and that the influence radius follows the empirical scaling relation \citep{StoneMetzger16}
\begin{equation}
r_{\rm infl} = 16~{\rm pc}~\left( \frac{M_\bullet}{10^8 M_\odot} \right)^{0.69}. \label{eq:SMInfl}
\end{equation}
The DF $f_\epsilon(\epsilon)$ is calculated numerically using Eddington's integral (considering both the stellar and the SMBH potential).  While Eddington's integral is strictly valid only for isotropic stellar systems, and overestimates $f_\epsilon(\epsilon)$ for radially biased systems, we find that the net effect on the TDE rate is small.  An example family of runs from this grid, with $\beta_0 = 0.5$, is shown in Fig. \ref{fig:anisoGalaxies}.  We see that galaxy-integrated rate enhancements $\mathcal{R}_\beta(t)\equiv \dot{N}_\beta(t) / \dot{N}_{\rm iso}(t)$ are the smallest, and wash out the fastest, for low mass galaxies.  This is almost entirely understandable in terms of the longer relaxation times in high mass galaxies: for a fixed physical post-starburst age $t$, a $M_\bullet = 10^8 M_\odot$ galactic nucleus will have elapsed much less dimensionless time $\tau$ than a galactic nucleus with $M_\bullet = 10^6 M_\odot$.  Of course, a countervailing effect is that the high mass SMBHs with the largest anisotropic rate enhancements are less common than their low mass counterparts; we return to this competition  in \S \ref{sec:DTD}. 

Interestingly, our hypothesis that most flux into the LC originates from orbital energies $\epsilon \approx \epsilon_{\rm crit}$ is incorrect.  LC flux is much less steeply peaked for highly anisotropic systems ($\beta > 0$) than it is for isotropic ones, and a large portion of TDEs in radially biased galactic nuclei originate from physical scales $r \gg r_{\rm infl}$.  The reason for this is that the ``full loss cone flux'' which sets an upper limit on TDE rates in isotropic galactic nuclei,
\begin{equation}
\mathcal{F}_{\rm full}^{\rm iso}(\epsilon){\rm d}\epsilon = 4\pi^2 f(\epsilon) J_{\rm LC}^2(\epsilon){\rm d}\epsilon, \label{eq:FFullIso}
\end{equation}
drops off steeply as one moves to energies $\epsilon < \epsilon_{\rm crit}$.  While an isotropic spherical system cannot produce LC fluxes greater than $\mathcal{F}_{\rm full}^{\rm iso}$, an anisotropic one can; its equivalent upper limit on LC flux is 
\begin{equation}
\mathcal{F}_{\rm full}(\epsilon){\rm d}\epsilon \approx 4\pi^2 f(\epsilon) J_{\rm LC}^2(\epsilon) j_{\rm LC}^{-2\beta}(\epsilon){\rm d}\epsilon \label{eq:FFull}
\end{equation}
provided the $f_{\rm j}(j)$ DF is described well by Eq. \ref{eq:ICs} down to values of $j$ near $j_{\rm LC}$.  For $\beta>0$, Eq. \ref{eq:FFull} declines less steeply for $\epsilon < \epsilon_{\rm crit}$ than does Eq. \ref{eq:FFullIso}, allowing stars from well beyond the critical radius to contribute significantly to the TDE rate.  This raises the possibility that resolved spectroscopy and Schwarzschild modeling of nearby E+A galaxies could directly test the premise of this solution to the PSP, but we defer a detailed investigation of this prospect for future work.

In summary, radial orbital biases among the stars deposited in galactic nuclei following a starburst can provide large enhancements to TDE rates (up to $\sim 100$), which wash out over an energy relaxation time.  Larger galaxies are more promising hosts for these rate enhancements because of their longer relaxation times.  For a stellar density profile $\rho(r) \propto r^{-\gamma}$, the radial anisotropy erodes from the inside out if $\gamma>3/2$, and from the outside in if $\gamma < 3/2$.  Regardless of the value of $\gamma$, the anisotropy among stars of {\it fixed energy} erodes from the inside out in $j$-space.  In observed TDE host galaxies, of age $t \gtrsim 10^8~{\rm yr}$, rate enhancements can be large enough to explain observations if large initial values of $\beta_0$ do not trigger the radial orbit instability.  We provide a more detailed comparison to observations in \S \ref{sec:DTD}.

\section{SMBH Binaries}
\label{sec:SMBHB}

Several studies have demonstrated that the dynamical presence of a SMBH binary in a galactic nucleus can enhance rates of tidal disruption by several orders of magnitude, relative to an otherwise similar nucleus with a single, stationary SMBH.  TDE rates are enhanced by the Kozai-Lidov mechanism \citep{Ivanov+05} and by chaotic three-body scatterings \citep{Chen+09}, though the latter effect appears to dominate \citep{Chen+11}.  Because E+A and post-starburst galaxies are often the result of galaxy mergers, a TDE rate enhancement driven by SMBH binarity offers a tempting explanation for the observed PSP \citep{Arcavi+14}.  

One potential issue with this explanation is that the duration of TDE enhancement by the SMBHB is very short, $\sim 10^{5-6}~{\rm yr}$.  Although the TDE rate temporarily increases once a SMBH binary forms, as the SMBH binary ejects or tidally disrupts most of the  stars within the influence radius of the primary SMBH, binary hardening stalls and the TDE rate plummets to levels far below that of a standard galactic nucleus \citep{Chen+08}.  By combining dynamical estimates of the number of TDEs per SMBHB merger with approximate galaxy merger rates, \citet{WeggBode11} estimate that SMBHBs contribute only $\sim 2\%$ of the total number TDEs over cosmic times, insufficient to explain the observed PSP.  However, given the many uncertainties in the calculation of \citet{WeggBode11}, this channel deserves further exploration.

In this section, we calculate the total TDE rate due to SMBHBs, in order to assess this scenario's overall viability as an explanation for the PSP.  Although the DTD itself is harder to quantify in this case than in the others we have considered thus far (\S \ref{sec:overdense}, \S \ref{sec:aniso}), we nevertheless estimate it in a limiting case.

\subsection{TDE Rate from SMBHBs}
\label{sec:SMBHBRates}

Estimating the SMBH contribution to the TDE rate requires several pieces of information from astrophysical modeling and observations, which we first review before describing the calculation itself:
\begin{enumerate}
\item The rate of galaxy mergers at redshift $z$ is estimated using the following fitting formula taken from the cosmological Illustris Simulation \citep{RodriguezGomez+15}:
\begin{align}
\frac{{\rm d}N}{{\rm d}\mathcal{Q}_\star{\rm d}t} =& A(z) \left(\frac{\mathcal{M}_\star}{10^{10}M_\odot} \right)^{\alpha (z)}\left[1+ \left(\frac{\mathcal{M}_\star}{2\times 10^{11}M_\odot} \right)^{\delta (z)}\right] \label{eq:mergerRate} \\
& \times \mathcal{Q}_\star^{\beta(z)+\gamma \log_{10}(\mathcal{M}_\star/(10^{10}M_\odot))}, \notag
\end{align}
where $\mathcal{M}_\star$ is the stellar mass of the larger galaxy and $\mathcal{Q}_\star<1$ is the (stellar) mass ratio of the merger.  The fitted constants and functions in this equation are compiled in Appendix \ref{app:mergerRate}.  
\item The SMBH mass $M_{\bullet}$ in each galaxy is calculated from the galactic bulge mass $\mathcal{M}_{\rm bulge}$ using the \citet{KormendyHo13} calibration of the $M_\bullet - \mathcal{M}_{\rm bulge}$ relationship, as
\begin{equation}
M_\bullet = 0.49^{+0.06}_{-0.05} \times 10^9 M_\odot \left(\frac{\mathcal{M}_{\rm bulge}}{10^{11}M_\odot} \right)^{1.16\pm0.08}. \label{eq:KH13}
\end{equation}
We estimate $\mathcal{M}_{\rm bulge}$ from the galaxy stellar mass $\mathcal{M}_\star$ using a bulge to total (B/T) relation from the mock galaxy catalog of \citet{vanVelzen17}, as provided in tabular form in Appendix \ref{app:mergerRate}.  The latter agrees qualitatively with similar work from the SDSS sample of \citet{Kim+16}.
\item If the smaller merger partner also contains a SMBH (with a mass also given by the $M_\bullet  - \mathcal{M}_{\rm bulge}$ relation and the above B/T prescription), then it will inspiral through dynamical friction to the center of the merger product, ultimately forming a SMBHB.  This inspiral occurs on the dynamical friction timescale, which we estimate as \citep{Taffoni+03}
\begin{equation}
T_{\rm DF} = A \frac{x_0^2 R_{\rm h}^2V_{\rm h}}{GM_{\rm s}} \label{eq:DF},
\end{equation}
where $R_{\rm h}$ is the virial radius of the primary galaxy's halo, $V_{\rm h}\equiv (GM_{\rm h}/R_{\rm h})^{1/2}$ is the circular orbital speed at that location, $x_0 \equiv R_0 / R_{\rm h}$ (where $R_0$ is the initial orbital separation), and $M_{\rm h}$ ($M_{\rm s}$) is the total halo mass of the primary (secondary) galaxy.  $A$ is a dimensionless constant defined in Appendix \ref{app:mergerRate}, which depends on the properties of the merging galaxies (e.g. central density profile, concentration parameter of the halo), and, more uncertainly, on the properties of the merger: both orbital circularity $\varepsilon\le1$ and initial separation $R_0$.  Because $R_{\rm h} \propto M_{\rm h}^{1/3}$, $T_{\rm DF} \propto A x_0^2/Q_{\rm h}$, where $Q_{\rm h} \equiv M_{\rm s} / M_{\rm h}$.  The extremely weak dependence of $A$ on $M_{\rm h}$ (see Appendix \ref{app:mergerRate}) renders $T_{\rm DF}$ almost completely independent of $M_{\rm h}$ when all other merger properties (e.g. $Q_{\rm h}$, $\varepsilon$, $x_0$) are held fixed.

Because the TDE rate is enhanced only near the end of the dynamical friction inspiral, $T_{\rm DF}$ provides an estimate of the delay time between the merger and the epoch of TDE rate enhancement.
\item Finally, the total number of TDEs when forming a bound SMBHB is estimated as \citep{LiuChen13}
\begin{equation}
N_{\rm TDE} \approx \frac{7\times 10^4}{(3-\gamma)^{1/2}} Q^{(2-\gamma)/(6-2\gamma)}\left( \frac{M_\bullet}{10^7M_\odot} \right)^{11/12},
\label{eq:NTDE}
\end{equation}
where $Q<1$ is the SMBHB mass ratio computed from the galaxy mass ratio $\mathcal{Q}_\star$ using Eq. \ref{eq:KH13}.  This analytic result is empirically calibrated from the three-body scattering simulations of \citet{Chen+11}, who computed TDE rates due to the perturbation of stellar orbits by the potential of a secondary SMBH in a galactic nucleus.  The scalings entering this expression roughly agree with the independent work of \citet{Wegg13}, although the latter simulations find a prefactor which is $\approx 5$ times smaller.  Our use of Eq. \ref{eq:NTDE} therefore places an upper limit on the number of SMBHB-catalyzed TDEs, as the true number could be smaller by a factor of a few (though we note that both \citealt{Chen+11} and \citealt{Wegg13} consider SMBHBs in spherical star clusters; it is not yet known whether nonspherical geometries can enhance the total number of SMBHB-induced disruptions).
\end{enumerate}

From the above information, we estimate the per-galaxy TDE rate at redshift $z$ as the integral of the galaxy merger rate (Eq.~\ref{eq:mergerRate}) multiplied by the number of TDEs per merger (Eq.~\ref{eq:NTDE}),
\begin{equation}
\dot{N}_{\rm SMBHB}(\mathcal{M}_\star, z)=\int_{\mathcal{Q}_{\rm min}(z)}^{1}\frac{{\rm d}N}{{\rm d}\mathcal{Q}_\star{\rm d}t}\Big\rvert_{z_{\rm DF}} N_{\rm TDE}(\mathcal{M}_\star, \mathcal{Q}_\star){\rm d}\mathcal{Q}_\star. \label{eq:NDotSMBHB}
\end{equation}
The lower limit of integration $0.01 \lesssim \mathcal{Q}_{\rm min}(z) \lesssim 0.1$ is set by the lowest merger mass ratio capable of producing a SMBHB within the lookback time to when the merger took place.  This is determined by equating the dynamical friction timescale (Eq.~\ref{eq:DF}) to the lookback time between the observing redshift $z$ and the redshift of the merger $z_{\rm DF}$, considering only mergers that took place at $z_{\rm DF}<3$.  The galaxy merger rate ${\rm d}N/{\rm d}\mathcal{Q}_\star{\rm d}t$ in Eq.~(\ref{eq:NDotSMBHB}) is evaluated at the unique merger redshift $z_{\rm DF}(\mathcal{M}_\star, \mathcal{Q}_\star, z)$ which yields a SMBHB by the event redshift $z$ (we compute $z_{\rm DF}$ by numerically solving Eq. \ref{eq:DF} for $z$).

\begin{figure}
\includegraphics[width=85mm]{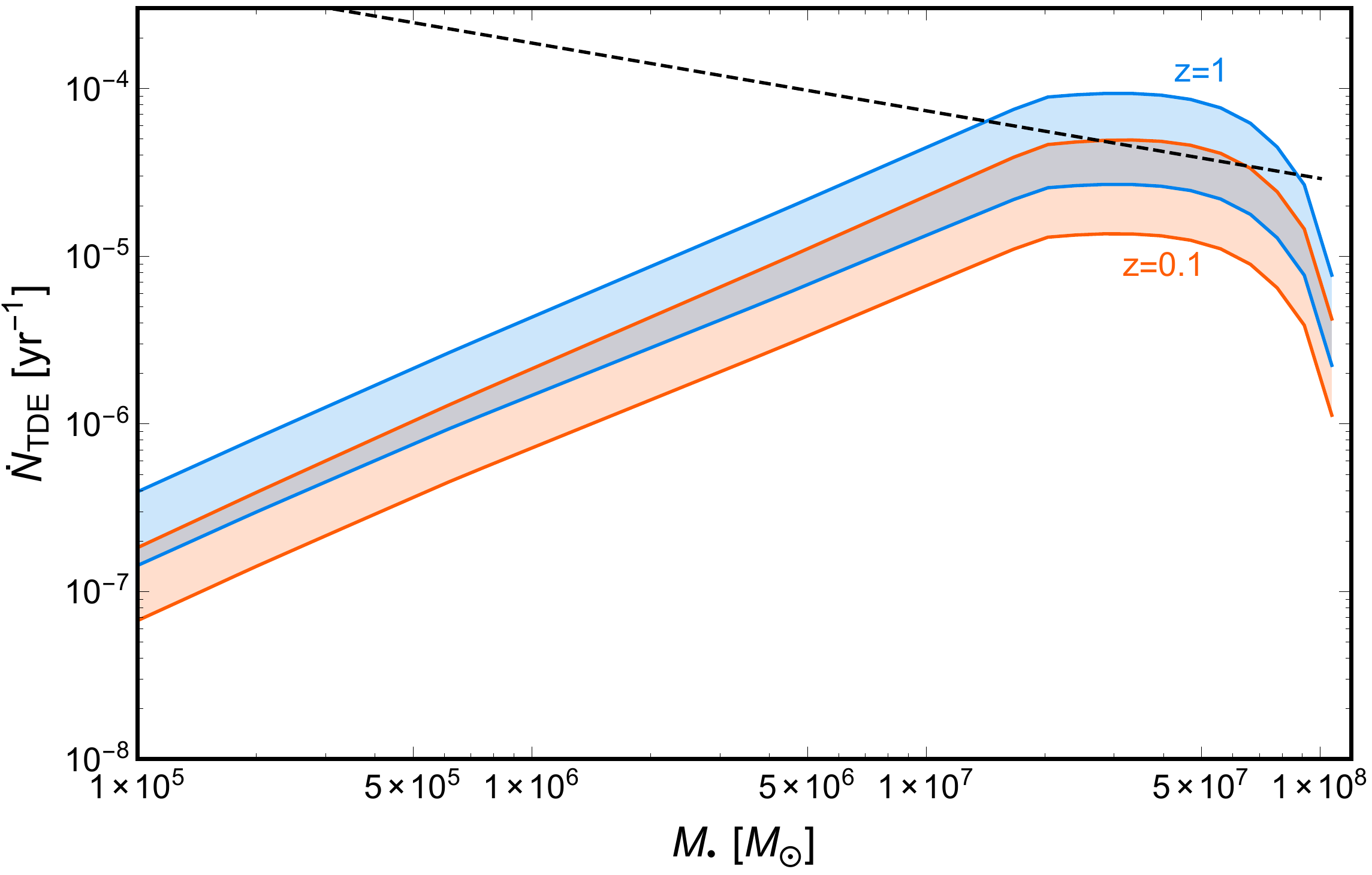}
\caption{Cosmically-averaged per-galaxy TDE rate $\dot{N}_{\rm SMBHB}$ of the binary SMBH channel, as a function of the mass $M_\bullet$ of the descendant SMBH.  Orange and blue lines correspond to the rate at redshifts $z=0.1$ and $z=1$, respectively.  Shaded regions denote the uncertainty range defined by adopting ``optimistic" versus ``pessimistic" values for uncertain parameters, such as the average circularity of the galaxy merger ($\epsilon=0.5$ or $0.9$), initial merger radius ($R_0=0.4R_{\rm h}$ or $0.8R_{\rm h}$), and slope of the density profile of the nuclear star cluster ($\gamma=1.9$ or $1.2$).  Averaged over the SMBH mass function, the local ($z=0.1$) rates are far below those predicted in {\it typical} (non-SMBH binary hosting) galactic nuclei through two-body interactions, though they are competitive at the high mass end ($M_\bullet \gtrsim 10^7 M_\odot$).  The best fit power law to the predicted two-body TDE rate of typical galaxies is shown as a black dashed line \citep{StoneMetzger16}.}
\label{fig:NDotSMBHB}
\end{figure}

Fig. \ref{fig:NDotSMBHB} shows the per-galaxy TDE rate $\dot{N}_{\rm TDE}$ as a function of the descendent SMBH mass $M_\bullet$ at two characteristic redshifts, $z = 0.1$ (orange lines) and $z = 1$ (blue lines).  One uncertainty in our calculation arises because we must make assumptions about the circularity $\varepsilon \le 1$ of the merger and the initial orbital radius $R_0$ of the satellite galaxy, which enter into the dynamical friction timescale.  Another uncertain parameter is the power law density slope $\gamma$ of the primary's nuclear star cluster.  The shaded error region in each curve brackets these uncertainties between an optimistic ($\varepsilon=0.5$, $R_0=0.4R_{\rm h}$, $\gamma=1.9$) and a pessimistic ($\varepsilon=0.9$, $R_0=0.8R_{\rm h}$, $\gamma=1.2$) case.  Shown for comparison is a power-law fit to the average TDE rate as a function of SMBH mass, calculated from two-body scattering based on measured stellar density profiles from a large population of nearby galaxies \citep{StoneMetzger16}.  

The rate of TDEs arising from SMBHBs increases with increasing (primary) SMBH mass $M_\bullet$, as massive galaxies experience a moderately greater number of mergers than smaller ones (Eq. \ref{eq:mergerRate}) and produce far more SMBHB-induced disruptions in each (Eq. \ref{eq:NTDE}).  
The TDE rate rolls over above the Hills mass\footnote{We compute population-averaged Hills masses following the prescription of \citet{StoneMetzger16}, i.e. assuming a Kroupa present-day mass function truncated above $M_\star = 1M_\odot$.} of $M_{\rm H}^{2/3} = R_\star c^2/(2GM_\star^{1/3})$, the maximum mass SMBH that can produce visible flares from tidal disruption \citep{Hills75}.  Since the vast majority of binary-induced TDEs are caused by the primary SMBH \citep{Wegg13}, we only consider its Hills mass in our computations.  

At $z=0.1$ the per-galaxy rate of SMBHB-catalyzed TDEs is generally less than the $\dot{N} \sim 10^{-4}~{\rm yr}^{-1}$ predicted by theory \citep{StoneMetzger16} and suggested by recent modeling of the observed TDE luminosity function \citep{vanVelzen17}.  However, the SMBHB-catalyzed TDE rate becomes competitive at the most massive end of TDE hosts ($M_\bullet \sim 10^{7.5} M_\odot$), and could under optimistic assumptions account for an order unity fraction of disruptions there.  The SMBHB TDE rate in our model grows with redshift, increasing by a factor of a few from $z=0.1$ to $z=1$.  However, because our model does not account for the uncertain redshift evolution of the $M_\bullet - \mathcal{M}_{\rm bulge}$ scaling relation, its predictions at high $z$ should be taken with caution.  That said, the redshift evolution of the SMBH mass function (cosmic downsizing) suggests that the TDE rate due to two-body relaxation could decline quickly with redshift, in which case the relative contribution of SMBHB TDEs may grow with $z$ \citep{Kochanek16}.  

By combining the per-galaxy TDE rate from SMBHBs with the local ($z\approx 0.1$) galaxy mass function of \citet{StoneMetzger16}, we estimate the volumetric TDE rate from the SMBHB channel to be $2.2\times 10^{-7}~{\rm Mpc}^{-3}~{\rm yr}^{-1} < \dot{n}<8.0\times 10^{-7}~{\rm Mpc}^{-3}~{\rm yr}^{-1}$.  While this range falls far below empirically-calibrated two-body TDE rate estimates of $\dot{n} \gtrsim 3.0\times10^{-6}~{\rm Mpc}^{-3}~{\rm yr}^{-1}$ \citep{StoneMetzger16}, it is comparable to or larger than some observationally estimated rates (for example, \citet{vanVelzenFarrar14} estimate a volumetric rate of $\dot{n}=4-8\times 10^{-8 \pm 0.4}~{\rm Mpc}^{-3}~{\rm yr}^{-1}$).  However, a recent analysis by \citet{vanVelzen17} finds a higher TDE rate of $\dot{n}\approx 7.4\times 10^{-7}~{\rm Mpc}^{-3}~{\rm yr}^{-1}$, which increases to $\dot{n}\approx 2.7\times 10^{-6}~{\rm Mpc}^{-3}~{\rm yr}^{-1}$ if the actual TDE luminosity function extends just a factor of a few lower, to a value of $\approx 10^{42}~{\rm erg}~{\rm s}^{-1}$, comparable to the dimmest observed flares.

If the broader TDE rate discrepancy indeed arises due to some combination of observational selection effects (e.g. dust extinction, spectroscopic followup choices in optical surveys, a wide TDE luminosity function), such that theoretical rate estimates are roughly correct, then we conclude that SMBHBs are relegated to a subdominant channel for TDE production.  They would in this case contribute at most $7.5-26\%$ of the ``normal rate" arising from two-body relaxation\footnote{The fraction of SMBHB-induced TDEs would be even lower if the occupation fraction of massive black holes with $M_\bullet \lesssim 10^6 M_\odot$ is high; this comparison assumes the conservative theoretical 2-body rate of $\dot{n} = 3.0\times10^{-6}~{\rm Mpc}^{-3}~{\rm yr}^{-1}$}, and are unlikely to explain the PSP.  Alternatively, if something is wrong with standard two-body relaxation calculations \citep{WangMerritt04, StoneMetzger16}, and lower observationally inferred rates are correct (\citealt{Donley+02, vanVelzenFarrar14}, but see also \citealt{Esquej+08, Holoien+16}), then TDEs sourced by SMBHBs may constitute an order unity fraction of the TDE sample, and thus may serve as viable explanations of the PSP.  Regardless, we have found a somewhat higher fractional rate of SMBHB-induced TDEs than the $\approx 3\%$ estimated by \citet{WeggBode11}, though at least part of this difference stems from the different three-body scattering results found by \citet{WeggBode11} versus those from \citet{Chen+11} used here.

Independent of these uncertainties on the total rate, one additional comparison to observations can be gleaned from Fig. \ref{fig:NDotSMBHB}~that appears to disfavor SMBHBs as a dominant TDE production channel.  As noted earlier, most TDEs sourced from SMBHBs arise from primary SMBHs of mass $10^7 M_\odot \lesssim M_\bullet \lesssim 10^8 M_\odot$, a result which is robust to the observing redshift and choice of free parameters in our dynamical friction model.  However, this result is in strong tension with the bottom-heavy SMBH mass distribution of TDE hosts found by \citep{Wevers+17}, based on a sample of twelve optically-selected TDEs.  It is in less, but still noticeable, tension with the luminosity-weighted TDE host mass function constructed from this sample \citep{vanVelzen17}.

\subsection{The SMBHB DTD}

The DTD for the SMBHB channel represents a convolution of two different delay times - the dynamical friction time $T_{\rm DF}$ (which determines the delay between the galaxy merger and the enhancement of the TDE rate due to SMBHB formation) and the delay between the galaxy merger and the starburst, $\Delta T_{\rm burst}$.  While the former can be estimated (albeit with some uncertainties) from Eq. \ref{eq:DF}, the time delay between merger and starburst is much more uncertain, and ultimately must be determined from cosmological simulations.

Nevertheless, using the tools assembled here, we can still estimate the DTD in a highly idealized version of this scenario, in which the merger-starburst delay is much smaller than the delay between merger and
SMBHB formation.  In this limit, the fraction of SMBHB-triggered TDEs that occur before a post-starburst age $t$ is given by
\begin{equation}
C(t; \mathcal{M}_\star, z)=\frac{\int_{\mathcal{Q}_{\rm min}(t, z)}^{1}\frac{{\rm d}N}{{\rm d}\mathcal{Q}_\star{\rm d}t}|_{z_{\rm DF}} N_{\rm TDE}(\mathcal{M}_\star, \mathcal{Q}_\star){\rm d}\mathcal{Q}_\star}{\dot{N}_{\rm SMBHB}(\mathcal{M}_\star, z)}. \label{eq:CSMBHB}
\end{equation}
Here we have defined a more general $\mathcal{Q}_{\rm min}(t, z)$ as the minimum galaxy mass ratio that will produce a SMBHB within a time $t<t_{\rm H}$ of the merger itself.

\begin{figure}
\includegraphics[width=85mm]{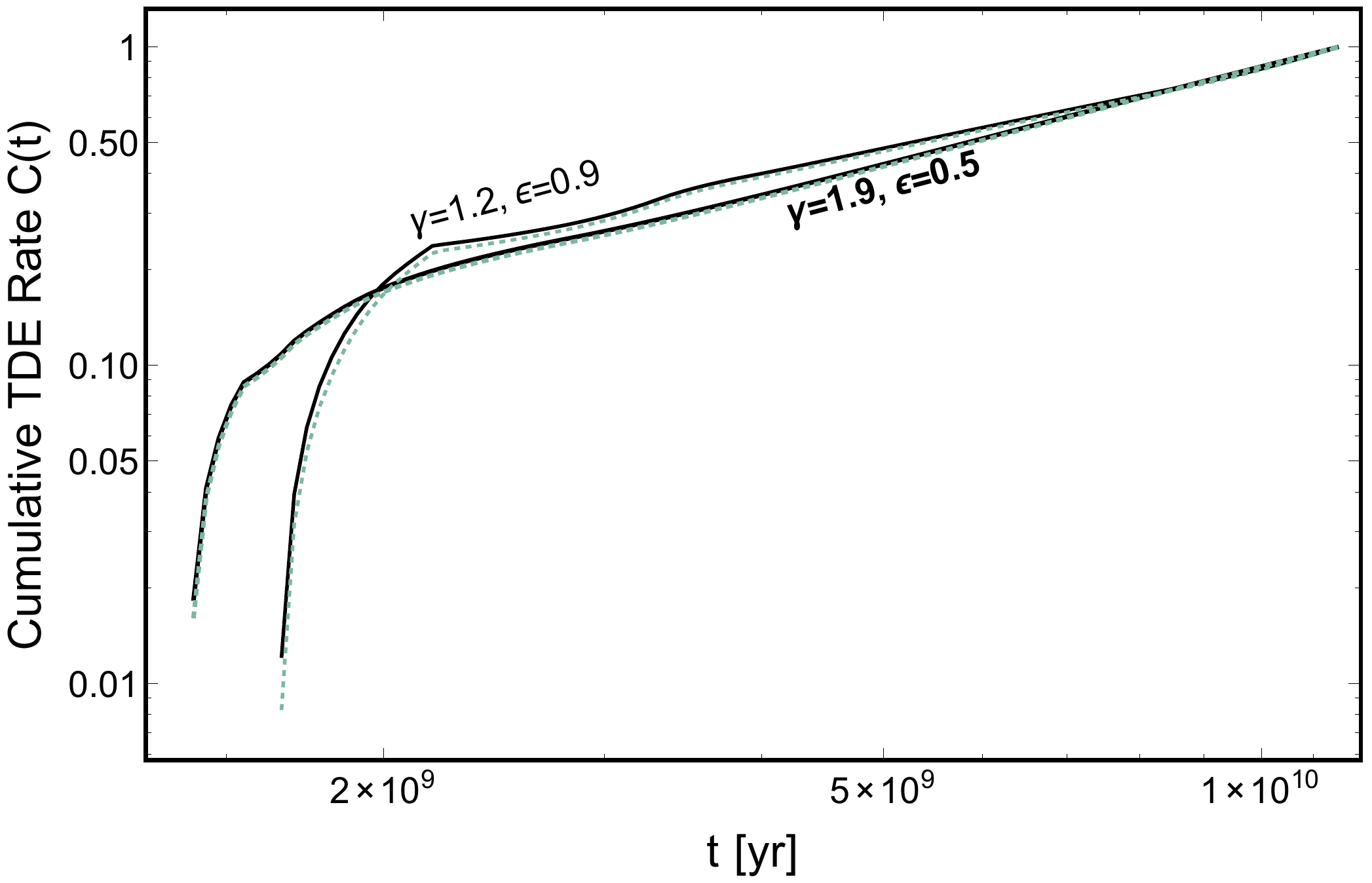}
\caption{Cumulative distribution $C(t)$ of the local ($z = 0.1$) TDE rate induced by SMBHBs as a function of the post-merger age $t$.  Thick/thin lines correspond, respectively, to optimistic/pessimistic choices of free parameters entering the SMBH rate calculation (see \S \ref{sec:SMBHBRates}).  Green dashed and black solid lines show results for primaries masses of $10^{6}M_{\odot}$ and $10^7 M_\odot)$, respectively.  The  strong similarities between these curves reflects how the primary mass scales out of Eq. \ref{eq:DF}, as explained in \S \ref{sec:SMBHBRates}.  The dearth of SMBHB-induced TDEs for post-merger ages $t<10^9~{\rm yr}$ disfavors the SMBHB explanation for the PSP, which requires a rate enhancement on timescales $t \lesssim 10^{8}$ yr.}
\label{fig:SMBHBCDTD}
\end{figure}

Our results for this cumulative distribution, which represents the {\it time integral} of the DTD, is shown in Fig.~\ref{fig:SMBHBCDTD}, shown separately for SMBHBs with primary masses of $M_\bullet = 10^{6},10^{7} M_\odot$.  We observe that essentially no SMBHB-triggered TDEs occur at early times $t<10^9~{\rm yr}$ after the galaxy merger.  Therefore, unless the average delay between a galaxy merger and its starburst is fine-tuned to occur in the narrow window $\Delta T_{\rm burst}  \approx (0.9-1) T_{\rm DF}$, there is no reason to expect a large fraction of TDEs should originate in galaxies with post-starburst ages $t\sim 10^8~{\rm yr}$.  Since the merger-starburst delay time is set by completely different physical processes than those that set $T_{\rm DF}$, this requisite fine-tuning provides an additional argument against the SMBHB explanation for the PSP.

\section{Comparison to Observations}
\label{sec:DTD}

The previous sections outlined three dynamical explanations for the PSP, for two of which (the overdensity and radial anisotropy hypotheses) we have estimated theoretical DTDs.  The DTD in the SMBHB case is difficult to estimate outside the framework of cosmological simulations, but as this scenario is disfavored on multiple grounds (insufficient overall rate, a top-heavy SMBH mass distribution, and required fine-tuning of the time delay $\Delta T_{\rm burst}$ between galaxy merger and starburst), we do not consider it further.  We begin this section by addressing the interaction between the post-starburst preference and a broader TDE rate dilemma (\S \ref{sec:ratesDilemma}).  Then, we compare our theoretical DTDs to the observational data collected by \citealt{French+17} (\S \ref{sec:realDTD}).  

\subsection{TDE Rates}
\label{sec:ratesDilemma}
Two-body relaxation estimates of TDE rates in nearby galaxies generally find that (i) the total TDE rate is dominated by those from the lowest mass SMBHs that exist with a high occupation fraction\footnote{But see also \citet{Brockamp+11}.}; (ii) under the assumption of a broad SMBH mass function, the average per-galaxy TDE rate is $\dot{N}\gtrsim 10^{-4}~{\rm gal}^{-1}~{\rm yr}^{-1}$ \citep{WangMerritt04, StoneMetzger16}.  While the distribution of SMBH masses in optically-selected TDE hosts appears consistent with the first prediction (Wevers et al. 2017), the second prediction may be in significant tension with observations.  The observed rate as inferred from X-ray-selected flares ranges from $\dot{N} = 9\times 10^{-6}~{\rm gal}^{-1}~{\rm yr}^{-1}$ \citep{Donley+02} to $\dot{N}=2\times 10^{-4}~{\rm gal}^{-1}~{\rm yr}^{-1}$ \citep{Esquej+08}.  Rates inferred from optically-selected flares are also quite uncertain; the ASAS-SN survey, for instance, estimates a TDE rate of $\dot{N}=2-17\times 10^{-5}~{\rm gal}^{-1}~{\rm yr}^{-1}$ \citep{Holoien+16}.  

An accurate rate estimate is challenging to obtain, due to uncertain survey selection effects; for example, in many time domain surveys, it is hard to quantify what criteria leads certain transients to receive the spectroscopic follow-up necessary to result in a TDE classification.  Dust extinction may lead to an underestimation of the true TDE rate; conversely, the present TDE sample may be contaminated at some level with TDE impostors, such as nuclear supernovae \citep{Saxton+16}, unusually variable AGN \citep{Komossa+15}, or exotic types of stellar collisions \citep{MetzgerStone17}.  Ideally, rates should be estimated using a complete, flux-limited sample, but such an approach is hindered by the complicated selection functions that enter into most time domain surveys.  Only the small number of SDSS-selected TDEs ($N=2$) represents a truly flux-complete (down to a limiting peak TDE luminosity) sample, from which one infers a low rate of $\dot{N}=0.2-4.7 \times 10^{-5}~{\rm gal}^{-1}~{\rm yr}^{-1}$ \citep{vanVelzenFarrar14}.  However, the true TDE rate is probably higher than this: recent modeling of the optical TDE luminosity function indicates that large, generally unseen populations of faint TDEs could be biasing observational inferences.  When the low detection rate of these faint TDEs is accounted for, one finds that they dominate the volumetric rate, and this yields a flux-corrected per-galaxy TDE rate $\dot{N}\approx 1\times 10^{-4}~{\rm gal}^{-1}~{\rm yr}^{-1}$ \citep{vanVelzen17}, potentially eliminating the tension with theoretical two-body rates.  

The overall magnitude of the TDE rate discrepancy in normal (non post-starburst) galaxies remains an open question, but one that is important for using DTDs to discriminate between different dynamical explanations of the PSP.  Because the bottom end of the TDE luminosity function remains poorly constrained, observational estimates of the PSP are most accurately expressed as {\it relative} rate enhancements $\mathcal{R}$ as a function of post-starburst age, rather than in terms of the {\it absolute} per-galaxy rate $\dot{N}$.  By making the reasonable assumption that the TDE luminosity function is the same in normal and post-starburst galaxies, this enables us to translate theoretical predictions of absolute TDE rates (e.g. \S \ref{sec:overdense}) into predictions about relative rates, which we compare to data in the next subsection.  However, we caution that this assumption (and thus our conclusions derived from it) could be invalid if the PSP arises due to non-dynamical effects, such as preferential dust obscuration in normal galaxies.

\subsection{The Delay Time Distribution}
\label{sec:realDTD}

\citet{French+16} found that 6 of 8 TDEs in their optical/UV-selected sample occurred in galaxies with strong Balmer absorption features, consistent with a starburst occuring $t_{\star} \sim$ 100 Myr to 3 Gyr
ago.  Three of these are classical E+A galaxies, consistent with a recent starburst ($t_{\star} \sim 100-200$ Myr), while another three show weaker Balmer absorption features, consistent with a slightly older starburst with $t_{\star}\sim 0.3-3$ Gyr (but see also \citealt{French+17} for caveats on the latter set of post-starburst ages).  Only two TDEs in the sample occur in galaxies with old stellar populations $t_{\star} \gtrsim 1$ Gyr, implying a much lower per-galaxy TDE rate in this population.  By comparing the number of TDEs in each bin of post-starburst age to the frequency of these galaxy ages in SDSS data, and under the assumption that the TDE luminosity function (after absorption/extinction) is constant across galaxy types, one finds that the TDE rate is enhanced in E+A and more weakly Balmer-strong galaxies by factors of $\mathcal{R}=190^{+115}_{-100}$ and $\mathcal{R}=33^{+7}_{-11}$, respectively.  

More recent work, accounting for possible selection effects and employing larger TDE candidate samples, has found smaller but still significant overrepresentations of post-starburst galaxies among TDE hosts \citep{LawSmith+17, Graur+17}.  In particular, \citet{Graur+17} used the same galaxy selection criteria as \citet{French+16}, and found TDE rate enhancements in E+A and more weakly Balmer-strong galaxies of $\mathcal{R}=36^{+22}_{-18}$ and $\mathcal{R}=18^{+8}_{-6}$, respectively.  However, the careful stellar population modeling of \citet{French+17} has found the most accurate post-starburst age estimates so far for TDE host galaxies, so we rely on this work in estimating the observational DTD.  We split their sample of eight TDE hosts into three categories.  First, two TDE hosts are not in post-starburst galaxies.  Second, five TDE hosts have a  ``post-starburst age'' (the age since the {\it end} of the burst) of $100-600~{\rm Myr}$.  Finally, one TDE host has a post-starburst age within $600-1100~{\rm Myr}$.  Under the assumption that these two equally sized age bins are equally represented in the $2.3\%$ of low-$z$ galaxies with significant H$\delta$ absorption \citep{French+16}, we estimate that for our ``young'' and ``old'' bins of post-starburst age, the rate enhancement factors are $\mathcal{R}=54^{+17}_{-20}$ and $\mathcal{R}=11^{+20}_{-9.0}$, respectively.  Here the upper and lower limits are one-$\sigma$ binomial confidence levels \citep{Gehrels86}.

\begin{figure*}
\includegraphics[width=170mm]{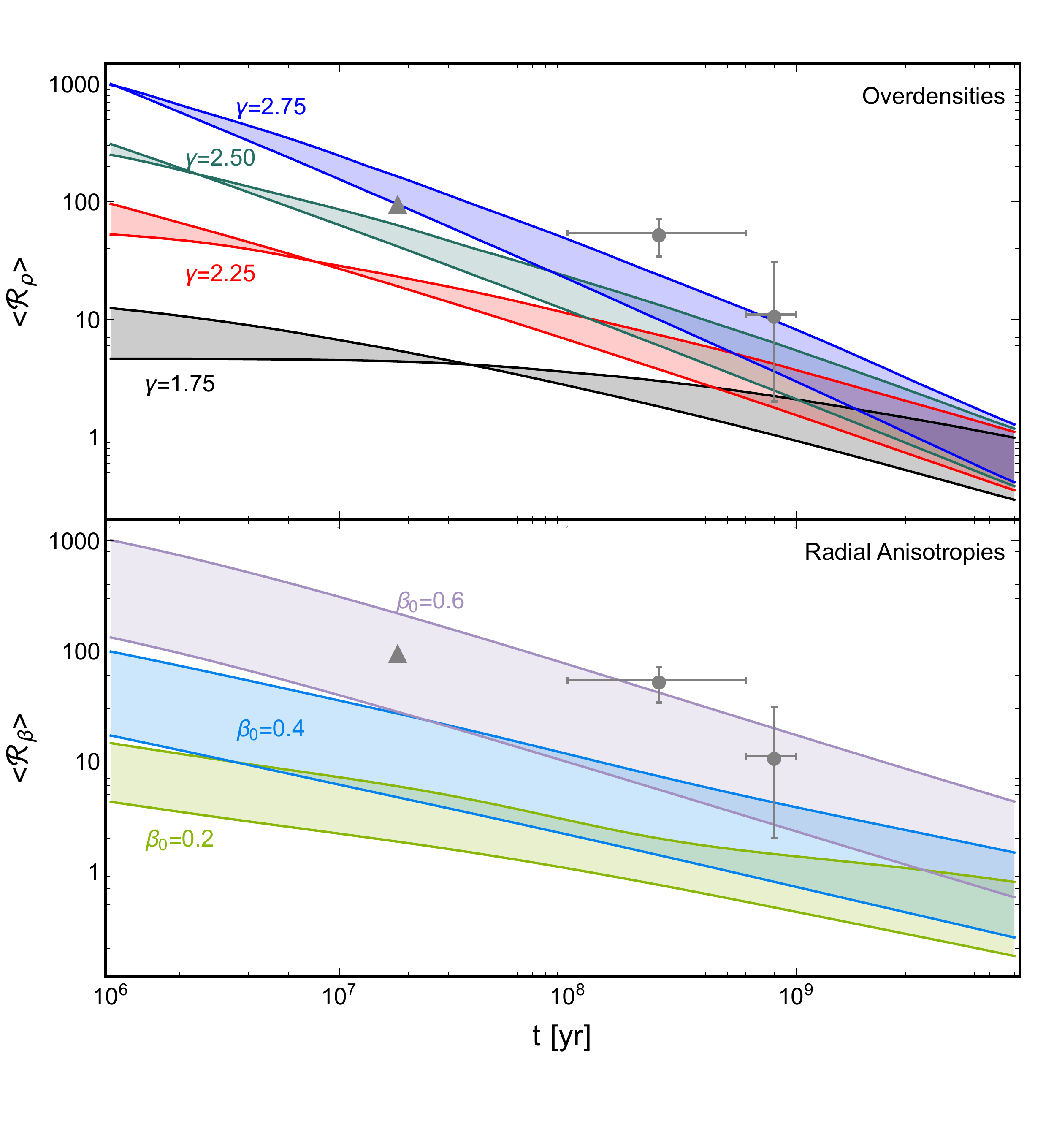}
\caption{Theoretical delay time distributions showing TDE rate enhancements as a function of stellar age $t$ of the nuclear star cluster.  The rate enhancements in the overdensity and radial anisotropy explanations of the post-starburst TDE preference are compared to observational data from \citet{French+17}, shown as gray circles.   As DTDs must be integrated over the SMBH mass function, the shaded regions denote the uncertainties arising from whether we assume SMBH mass function to cut off below $M_{\bullet} = 10^{5}M_{\odot}$ or $M_{\bullet} = 10^{6}M_{\odot}$.  A gray triangle shows the tentative TDE rate enhancement in starbursting galaxies, as inferred by the detection of a possible TDE in a ULIRG by \citealt{Tadhunter+17} (though we caution that the TDE interpretation of this single event requires verification).  {\it Top panel:} Rate enhancements $\langle \mathcal{R}_\rho \rangle$ for overdense post-starburst nuclei are shown.  The black, red, green, and blue curves show the evolution of initial power law slopes $\gamma=1.75$, $\gamma=2.25$, $\gamma=2.5$, and $\gamma=2.75$, respectively.  Each set of initial conditions has been given an initial normalization such that all Fokker-Planck simulations reproduce observed influence radii (Eq. \ref{eq:SMInfl}) after $10^{10}~{\rm yr}$ of evolution.  Observed post-starburst rate enhancements can be reproduced if $\gamma \gtrsim 2.5$.  {\it Bottom panel:} Rate enhancements $\langle \mathcal{R}_\beta \rangle$ for radially anisotropic post-starburst nuclei are shown.  The lavender, cyan, and light green curves show time evolution of initial anisotropies $\beta_0=0.6$, $\beta_0=0.4$, and $\beta_0=0.2$, respectively.  Observed post-starburst rate enhancements can be reproduced if $\beta_0 \gtrsim 0.5$, which is close to the onset threshold of the radial orbit instability.}
\label{fig:DTDDataAll}
\end{figure*}

A tidal disruption flare candidate was recently discovered \citep{Tadhunter+17, Dou+17} in a starbursting ultra-luminous infrared galaxy (ULIRG).  Although further observations are needed to substantiate the TDE origin of this event, its confirmation would suggest that TDE rates are even higher in actively starbursting galaxies, $\gtrsim 10^{-2}~{\rm yr}^{-1}$.  If the TDE rate in normal galaxies is $\dot{N} \sim 10^{-4}~{\rm gal}^{-1}~{\rm yr}^{-1}$, this implies a rate enhancement of $\mathcal{R} \gtrsim 100$ in starburst galaxies ($t_\star \sim 10^7~{\rm yr}$).  Given the still uncertain nature of the ULIRG flare, we do not focus on explaining this intriguing data point, but still present it for comparison.

In order to compare the observed rate enhancement to our theoretical predictions of the overdensity (\S \ref{sec:overdense}) and radial anisotropy (\S \ref{sec:aniso}) hypotheses, we must define $M_\bullet$-averaged rate enhancement factors $\langle \mathcal{R}_{\rho}(t) \rangle$ and $\langle\mathcal{R}_{\beta}(t) \rangle$, respectively.  We do not compute an averaged rate enhancement for the SMBHB hypothesis (\S \ref{sec:SMBHB}) because, as already discussed, this scenario cannot explain the PSP absent fine-tuning.  These mass-averaged relative DTDs are calculated by integrating our earlier, mass-dependent DTDs $\dot{N}_\rho(t, M_\bullet)$ and $\dot{N}_\beta(t, M_\bullet)$ over the SMBH mass function $\phi(M_{\bullet})$ according to 
\begin{align}
\langle \mathcal{R}_{\rho}(t) \rangle =& \dot{n}_{\rm SM16}^{-1}\int_{M_{\rm min}}^{M_{\rm max}} \dot{N}_{\rho}(t, M_\bullet) \phi(M_\bullet){\rm d}\log_{10}(M_\bullet) \label{eq:rhoDTD} \\
\langle \mathcal{R}_{\beta}(t) \rangle =& \dot{n}_{\rm SM16}^{-1}\int_{M_{\rm min}}^{M_{\rm max}} \dot{N}_\beta(t, M_\bullet) \phi(M_\bullet){\rm d}\log_{10} (M_\bullet) \label{eq:betaDTD} \\
\dot{n}_{\rm SM16} =& \int_{M_{\rm min}}^{M_{\rm max}} \dot{N}_{\rm SM16}(M_\bullet) \phi(M_\bullet){\rm d}\log_{10} (M_\bullet),
\end{align}
where we take an empirical $\phi(M_\bullet)$ from \citealt{Shankar+09} (units of ${\rm Mpc}^{-3}~{\rm dex}^{-1}$).  Specifically, we use their tabulated low-redshift ($z=0.02$) mass function; as a check of robustness, we have compared it to a different $\phi(M_\bullet)$ derived from galaxy scaling relations \citep{StoneMetzger16} and found only modest changes to $\langle \mathcal{R}_{\rho}(t) \rangle$ and $\langle \mathcal{R}_{\beta}(t) \rangle$.  As the limits of integration, we adopt fiducial values for the minimum and maximum SMBH mass of $M_{\rm min} = 10^6 M_\odot$ and $M_{\rm max} = 10^8 M_\odot$, respectively.  Eqs. \ref{eq:rhoDTD} and \ref{eq:betaDTD} are both normalized by a mass-integrated volumetric TDE rate $\dot{n}_{\rm SM16}$, which we calculate using the best-fit per-galaxy TDE rate of \citet{StoneMetzger16}, 
\begin{equation}
\dot{N}_{\rm SM16}(M_\bullet) = 2.9 \times 10^{-5}~{\rm gal}^{-1}~{\rm yr}^{-1} \left(\frac{M_\bullet}{10^8 M_\odot} \right)^{-0.404},
\end{equation}
Fig. \ref{fig:DTDDataAll} compares the observed DTD to our mass-integrated theoretical predictions.  The top panel shows the relative rate enhancement $\langle \mathcal{R}_\rho(t)\rangle $ in the overdensity hypothesis (\S \ref{sec:overdense}) for different initial power law slopes $\gamma$ and a range of minimum SMBH masses $M_{\rm min}$.  The initial influence radii $r_{\rm infl,0}$ (or, equivalently, $\rho_1(0)$ values) are chosen so that the influence radius after $10^{10}~{\rm yr}$ of relaxation is equal to Eq. \ref{eq:SMInfl}, yielding a standard galactic nucleus\footnote{In order to set realistic initial conditions, we ran an exploratory grid of \textsc{PhaseFlow} models with a range of $\gamma$, $\rho_1(0)$ and $M_\bullet$ values, and measured the final influence radii $r_{\rm infl}$.  These models were flattened to a $\rho \propto r^{-1/2}$ core inside a collision radius $\sim 10^{-2}~{\rm pc}$ set by Eq. \ref{eq:tColl}, but our results are not sensitive to this flattening.  We then interpolated across $\rho_1(0)$ (for fixed $\gamma$, $M_\bullet$) to estimate the $\rho_1(0)$ value that will yield Eq. \ref{eq:SMInfl}, and used this value of $\rho_1(0)$ in a second grid of models across $\gamma$ and $M_\bullet$.  This second grid produced the DTDs used in Fig. \ref{fig:DTDDataAll}.  We find that an initial condition $\rho_1(0) = 0.617^\gamma \times 10^{6.32}~M_\odot~{\rm pc}^{-3} (M_\bullet / 10^6 M_\odot)^{-1.05+0.425\gamma}$ yields a final influence radius within a factor of two of Eq. \ref{eq:SMInfl}.}.  This approach yields a realistic $\langle \mathcal{R}_\rho(10^{10}~{\rm yr})\rangle \approx 1$.  Initial power law slopes $\gamma \gtrsim 2.5$ are capable of matching both post-starburst data points as well as the (very approximate) ULIRG data point.

Likewise, the bottom panel of Fig. \ref{fig:DTDDataAll} compares the observed DTD to the theoretical relative rate enhancement $\langle \mathcal{R}_\beta(t)\rangle$ in our anisotropic hypothesis (\S \ref{sec:aniso}) for different initial anisotropies $\beta_0$ and a range of minimum SMBH masses $M_{\rm min}$.  Lower values of $M_{\rm min}$ always reduce $\langle \mathcal{R}_\beta\rangle$ at fixed post-starburst age $t$, because lower mass SMBHs have, on average, shorter relaxation times, in which case initial anisotropies wash out more quickly.  As described earlier, larger values of $\beta_0$ lead to higher enhancements $\langle \mathcal{R}_\beta\rangle$.  The anisotropic DTDs can match the observed rate enhancements for $\beta_0 \gtrsim 0.5$.  Such large anisotropies lie close to the threshold for the radial orbit instability, and the viability of this hypothesis may depend on the nonlinear outcome of the ROI.  

Most TDEs produced by overdense galactic nuclei come from low-mass SMBHs.  The anisotropy hypothesis, on the other hand, is biased towards higher-mass SMBHs where relaxation times are long (see Fig. \ref{fig:anisoGalaxies}).  Roughly half of the rate enhancement $\langle \mathcal{R}_\beta(t)\rangle$ in the bottom panel of Fig. \ref{fig:DTDDataAll} comes from SMBHs in the mass range $10^7 < M_\bullet /M_\odot < 10^8$.  In this section we have cut off our integrals at a $M_{\rm max}=1\times 10^8 M_\odot$, in an approximation of the Hills mass.  While the anisotropic explanation of the PSP does not favor as top-heavy a distribution of TDE hosts as the SMBHB scenario does (Fig. \ref{fig:NDotSMBHB}), more accurate modeling of the Hills mass, and perhaps SMBH spin distributions \citep{Kesden12}, may be necessary for future work to compare predicted SMBH mass distributions to observations \citep{Wevers+17}.

\section{Conclusions}
\label{sec:conclusions}

We have developed a new tool, the TDE {\it delay time
  distribution}, for studying the unusual host galaxy preferences of TDEs.  This observable, already widely used in studies
of other transients such as gamma ray bursts and Type Ia supernovae, may in the near
future help discriminate between different explanations for the
post-starburst preference of observed TDE flares.  Most of these explanations are motivated by exotic stellar dynamical scenarios in the nuclei of post-starburst galaxies, and these scenarios generally have uncertain free parameters.  DTDs are useful not only for discriminating between explanations of the PSP, but also for parameter extraction in the context of an individual dynamical hypothesis.

We have also investigated a new hypothesis for the PSP based on the assumption that starbursts in nuclear star clusters place stars onto preferentially radial orbits around the SMBH.  Such radial orbits allow stars to diffuse
rapidly into the loss cone, enhancing the TDE rate over that in an otherwise
identical galactic nucleus with an isotropic stellar velocity field (to which the cluster relaxes at late times).
By numerically solving the 1D (angular momentum) Fokker-Planck
equation, we find that radial orbit anisotropies can enhance volumetric TDE rates by up to a factor of $\langle \mathcal{R}_\beta \rangle \approx 10-100$ (depending on the uncertain onset threshold of the radial orbit instability) on timescales of $\sim 10^{8}$ yr post-starburst.  This enhancement, which declines approximately as a power law in time in each galaxy ($\dot{N} \propto t^{-\beta_0}$), provides a promising explanation for the PSP.

The overdensity hypothesis, proposed theoretically in
\citet{StoneMetzger16} and investigated empirically in
\citet{StonevanVelzen16}, also appears capable of producing the large
per-galaxy TDE rates observed in post-starburst galaxies.  If the starburst produces an ultrasteep stellar cusp (initial power-law slope $\gamma \ge 9/4$) with a high density normalization, we find that large per-galaxy TDE rates of $\sim
10^{-3}~{\rm yr}^{-1}$ are possible in nuclei with young stellar populations of age $\sim
10^8~{\rm yr}$.  The TDE rate declines steeply with time following the starburst, approximately as $\dot{N} \propto t^{-(4\gamma-9)/(2\gamma-3)} / \ln t$.  When we choose initial density profile normalizations that, after $10^{10}~{\rm yr}$ of collisional evolution, reproduce observed nuclear properties (specifically, $r_{\rm infl}$), we find that stellar profiles with initial power-law slopes $\gamma \gtrsim 2.5$ can reasonably reproduce the DTD implied by current data.

By contrast, the original hypothesis of \citet{Arcavi+14} - that
the PSP may arise from a hidden population of SMBHBs - appears to contribute at most in a subdominant way to the observed rate enhancement.  While a detailed delay time distribution for the SMBHB rate enhancement is beyond the scope of this paper, and likely
can only be constructed from cosmological simulations, we have
generated a simplified DTD for this scenario (Fig.~\ref{fig:SMBHBCDTD}).  The SMBHB scenario is disfavored because dynamical friction sets a delay between galaxy merger and the formation of a hard SMBHB (i.e, the short-lived phase of TDE rate enhancement) that is typically $T_{\rm DF} \gtrsim 1 \times 10^9~{\rm yr}$.  Unless the delay between galaxy merger and starburst, which is set by an entirely different set of hydrodynamical processes, is fine-tuned to be $\lesssim 10 \%$ shorter than $T_{\rm DF}$, then the SMBHB scenario cannot explain the PSP.  Furthermore, by computing for the first time the distribution of primary SMBH masses in SMBHB-triggered TDEs, we have found a top-heavy distribution which is at odds with the bottom-heavy mass function of the observed TDE flare sample \citep{Wevers+17}.

While this paper provides a pioneering theoretical exploration of the TDE DTD, much additional work is needed to better characterize the time evolution of TDE rates.  First, our
investigations of the overdensity and radial anisotropy scenarios employed idealized 1D Fokker-Planck equations to study evolution in energy
and angular momentum space, respectively.  These calculations
could be improved by solving the 2D Fokker-Planck equation, with the Monte Carlo method \citep[e.g][]{DuncanShapiro83, FreitagBenz02, Vasiliev15}, or at much greater computational cost with full N-body simulations \citep[e.g.][]{Brockamp+11, Wang+16}.  The latter two approaches could also help explore TDE rates in post-ROI systems born with high $\beta_0$.  Second, we have not attempted to compute the DTDs for more
dynamically complex explanations of the PSP (e.g. a strongly triaxial
geometry for post-starburst galactic nuclei, or secular dynamical processes in eccentric stellar disks).  Third, we have largely ignored non-dynamical explanations of the PSP, such as preferentially low column depths in post-starburst galactic nuclei.  If normal galaxies generally suffer from high levels of nuclear dust extinction (photoelectric absorption from neutral gas), the detectability of their optically (X-ray) bright TDEs will fall in contrast to less obscured galaxy subpopulations.

Finally, we note that the three different dynamical hypotheses in this
paper - as well as other explanations of the PSP - may have another
potential discriminant: the distribution of TDE impact parameters
$\beta_{\rm TDE} = R_{\rm t}/R_{\rm p}$.  TDEs originating in SMBH
binaries, or in triaxial stellar potentials, feed stars into the loss
cone in the pinhole regime of disruption, in which a sample of
TDEs will have an impact parameter distribution $N(\beta_{\rm TDE}) \propto
1/\beta_{\rm TDE}$.  Conversely, the ultrasteep version of the overdensity
hypothesis, which appears in some ways to be the most promising
solution to the PSP, will be in the diffusive regime of
disruption, in which most TDEs have $\beta_{\rm TDE} \approx 1$ (though a minority of high $\beta_{\rm TDE}$ events will be supplied by strong scatterings - see \citealt{WeissbeinSari17}).  In the radial anisotropy scenario, the $\beta_{\rm TDE}$ distribution will depend on the details of the stellar density profile $\rho(r)$ and the initial anisotropy $\beta_0$, but will generally favor the pinhole regime.

Our current knowledge of the PSP is limited by small number statistics: existing observations
provide population ages for only a few dozen TDE host galaxies \citep{French+16, Graur+17}, and in many cases these ages are quite approximate \citep{French+17}.  However, the sample of TDEs is expanding rapidly and will continue to do so as more time domain surveys, such as {\it ZTF} and the {\it LSST}, come online.  As the stellar populations of more TDE hosts are characterized spectroscopically, it will become possible to construct
empirical delay time distributions that distinguish increasingly subtle variations between the theoretical DTDs computed here.  This will provide a straightforward test of different theoretical explanations for the peculiar host galaxy preference of TDEs, and other peculiar features of TDE demographics not yet identified.

\section*{Acknowledgments}
We gratefully acknowledge the assistance of Decker French in constructing the empirical DTD used in this paper, as well as fruitful discussions with Vicente Rodriguez-Gomez, Gregory Snyder, Paul Torrey, Sjoert van Velzen, and Ann Zabludoff.  NCS received financial support from NASA through Einstein Postdoctoral Fellowship Award Number PF5-160145, and thanks the Aspen Center for Physics for its hospitality during the completion of this work.  BDM and AG acknowledge support from NSF Astronomy and Astrophysics grants AST-1410950, AST-1615084; NASA Astrophysics Theory Program grants NNX16AB30G, NNX17AK43G; and Hubble Space Telescope Grant HST-GO-14785.004-A.  EV acknowledges support from the European Research council under the 7th Framework programme (Grant 321067).



\appendix

\section{Loss Cone Theory}
\label{app:LC}
Near a SMBH, stars are disrupted when angular momentum diffusion brings them onto radial orbits.  Diffusion through energy space contributes to TDE rates at a vastly lower level.  The loss cone is the region of phase space where the specific angular momentum obeys $J < J_{\rm LC}$, where $J_{\rm LC}^2\approx 2GM_\bullet r_{\rm t}$ is the angular momentum of the loss cone.  The reader interested in a thorough introduction to LC physics should consult \citet{CohnKulsrud78} and \citet{Merritt13}; our goal here is limited to summarizing those results relevant for our calculations in \S \ref{sec:overdense} and \S \ref{sec:aniso}.

We limit ourselves in this appendix to the special case where the stellar density profile obeys $\rho_{\star}(r) = \rho_0 (r/r_0)^{-\gamma}$ and $r \ll r_{\rm infl}$.  Inside the radius of influence, the Kepler potential of the SMBH is $\psi=GM_\bullet/r$; here we use the stellar dynamical convention where specific orbital energy $\epsilon$ is positive for bound orbits, as is the gravitational potential.  The one-dimensional velocity dispersion is
\begin{equation}
\sigma^2 = \frac{GM_\bullet}{r(1+\gamma)}.
\end{equation}
Assuming isotropic velocities, the distribution function is given by an Eddington integral:
\begin{equation}
f(\epsilon) = 8^{-1/2}\pi^{-3/2}\frac{\Gamma(\gamma+1)}{\Gamma(\gamma-1/2)} \frac{\rho_0}{\langle m_\star \rangle} \left(\frac{GM_\bullet}{r_0} \right)^{-\gamma}\epsilon^{\gamma-3/2}, \label{eq:DFEnergy}
\end{equation}
where $\Gamma$ is the standard Gamma function.  In \S \ref{sec:overdense}, we compute the relaxation time $t_{\rm r}$, which is defined in terms of a local diffusion coefficient for the parallel velocity component \citep{Merritt13}
\begin{align}
\langle (\Delta v_\parallel)^2\rangle=&\frac{32\pi^2}{3}G^2\langle m_\star^2\rangle \ln\Lambda v(F_4(v)+E_1(v)),\notag \\
E_1(v)=&\int_v^\infty \frac{v_{\rm f}}{v}f(v_{\rm f}){\rm d}v_{\rm f} \label{eq:vPar} \\
F_4(v)=&\int_0^v \left( \frac{v_{\rm f}}{v}\right)^4 f(v_{\rm f}){\rm d}v_{\rm f}. \notag
\end{align}
Here we have rewritten Eq. \ref{eq:DFEnergy} in terms of the local velocity $v$ with the substitution $\epsilon = \psi-v^2/2$.  We evaluate the diffusion coefficient at $v=\sqrt{3}\sigma(r)$.  In standard galactic nuclei, the use of $\langle (\Delta v_\parallel)^2\rangle$ usually gives a reasonable estimate of the energy relaxation time, but formally speaking, energy relaxation is governed by a nonlocal orbit-averaged diffusion coefficient.  When $\gamma \ge 2$, energy diffusion becomes strongly nonlocal.  We find numerically that using Eq. \ref{eq:vPar} to compute the normalization of the energy relaxation time $t_{\rm r}$ gives results off by a factor of a few when applied to the ultrasteep regime of \S \ref{sec:overdense}.

Deep in the Kepler potential of an SMBH, the orbit-averaged angular momentum diffusion coefficient (which arises from $\langle(\Delta v_\perp)^2 \rangle$, not Eq. \ref{eq:vPar}) can be approximately evaluated in closed form \citep[Appendix A]{StoneMetzger16}.  This is useful when considering the DTD of ultrasteep stellar density cusps, and in \S \ref{sec:overdense} we make use of the closed form expression for LC flux,
\begin{align}
\mathcal{F}(\epsilon) \approx & \frac{32\pi}{3\sqrt{2}}G^5 M_\bullet^3 \rho_0^2 \ln \Lambda \left( \frac{GM_\bullet}{r_0} \right)^{-2\gamma} \left( \frac{\gamma(\gamma-1/2)\Gamma(\gamma)}{\Gamma(\gamma+1/2)} \right)^2 \\
& \times \frac{3\tilde{Q}_{1/2}-\tilde{Q}_{3/2}+2\tilde{Q}_0}{\ln(GM_\bullet / (4\epsilon r_{\rm t}))} \epsilon^{2\gamma - 11/2}. \notag
\end{align}
Here the dimensionless terms
\begin{align}
\tilde{Q}_0 \approx & \frac{5\pi}{8(2\gamma -1)} \\
\tilde{Q}_{1/2} \approx & \pi^{1/2} \Bigg(\frac{1811-798\gamma+16\gamma^2}{120}\frac{\Gamma(4-\gamma)}{\Gamma(15/2-\gamma)}  \\
& + \frac{2\gamma - 1}{4(5-\gamma)(4-\gamma)} \frac{\Gamma(\gamma+1/2)}{\Gamma(\gamma+1)} \Bigg) \notag \\
\tilde{Q}_{3/2} \approx& \frac{\pi}{40\Gamma(\gamma-3)} \Bigg(\frac{\pi^{1/2}(8\gamma^2+118\gamma-325)\csc(\pi \gamma)}{\Gamma(15/2-\gamma)} \\
& - 15 \times \frac{2^{5-2\gamma}(1-2\gamma)^2 (2\gamma-7) (2\gamma - 5) (2\gamma - 3) \Gamma(2\gamma-8)}{\Gamma(\gamma+2)} \Bigg) \notag.
\end{align}
Note that the function $\tilde{Q}\equiv 3\tilde{Q}_{1/2}-\tilde{Q}_{3/2}+2\tilde{Q}_0$ has singularities at all integer and half-integer values of $\gamma$.  As these are removable singularities, they do not pose a serious challenge for numerical evaluation, but for the convenience of the reader, the following fitting formula is accurate to within $10\%$ for $0.5 \le \gamma < 2.95$: 
\begin{equation}
\tilde{Q} \approx \frac{1.99}{\gamma-1/2}-0.0657+0.597(\gamma-1/2)-0.192(\gamma-1/2)^2,
\end{equation}
where $\tilde{Q}\equiv 3\tilde{Q}_{1/2}-\tilde{Q}_{3/2}+2\tilde{Q}_0$.

In ultrasteep density cusps, $\dot{N}$ diverges as one integrates $\mathcal{F}(\epsilon)$ to more and more tightly bound orbits; if there is some large value of $\epsilon'$ above which the power law flattens, then $\dot{N} \sim \epsilon' \mathcal{F}(\epsilon')$.

\section{Prescriptions for Calculation of SMBHB TDE Rates}
\label{app:mergerRate}
\citet{RodriguezGomez+15} use the Illustris simulation to explore the merger rates of galaxies across a range of redshifts and mass ratios.  We employ their fitting formula, Eq. \ref{eq:mergerRate} in \S \ref{sec:SMBHB}, and list in this appendix the relevant ancillary fitting formulas and fitted constants.  In particular, $A(z) = A_0 (1+z)^\eta$, $\alpha(z) = \alpha_0(1+z)^{\alpha_1}$, $\beta(z) = \beta_0(1+z)^{\beta_1}$, and $\delta(z)=\delta_0(1+z)^{\delta_1}$.  $A_0 = 10^{-2.2287}~{\rm Gyr}^{-1}$, and the dimensionless fitted constants are $\eta = 2.4644$, $\alpha_0 = 0.2241$, $\alpha_1 = -1.1759$, $\beta_0 = -1.2595$, $\beta_1 = 0.0611$, $\gamma = -0.0477$, $\delta_0 = 0.7668$, and $\delta_1 = -0.4695$.

This formula is used to estimate merger rates between galaxies with total descendant stellar mass $\mathcal{M}_\star$, and mass ratio $\mathcal{Q}_\star$.  However, in early stages of the merger, dark matter is the dominant constituent powering the action of dynamical friction on the infalling satellite.  Thus, in order to compute the dynamical friction inspiral time, we need a stellar to halo mass relation (SHMR).  We employ the parametrized SHMR of \citet{Moster+13}, which was computed using abundance matching:
\begin{equation}
\frac{\mathcal{M}_\star}{M_{\rm h}} = 2N(z)\left[ \left( \frac{M_{\rm h}}{M_1(z)} \right)^{-B(z)} +  \left( \frac{M_{\rm h}}{M_1(z)} \right)^{C(z)}  \right]^{-1},
\end{equation}
where $M_{\rm h}$ is total halo mass.  This formula employs the following auxiliary fitting functions:
\begin{align}
\log_{10}M_1(z) = & M_{10}+M_{11}\frac{z}{z+1} \\
N(z) = & N_{10}+N_{11}\frac{z}{z+1} \\
B(z) = & B_{10}+B_{11}\frac{z}{z+1} \\
C(z) = & C_{10}+C_{11}\frac{z}{z+1},
\end{align}
where $M_{10}=11.590$, $M_{11}=1.195$, $N_{10}=0.0351$, $N_{11}=-0.0247$, $B_{10}=1.376$, $B_{11}=-0.826$, $C_{10}=0.608$, and $C_{11}=0.329$.

Likewise, we convert total stellar mass $\mathcal{M}_\star$ into bulge mass $\mathcal{M}_{\rm bulge}$ by using tabulated estimates for B/T ratios taken from a large sample of SDSS galaxies \citep{vanVelzen17}.  These tabulated estimates are binned by total stellar mass, and we linearly interpolate between different bins.  The results are qualitatively similar to parallel SDSS analysis in \citet{Kim+16}.  The precise tabulated data (van Velzen, private communication) is shown in Table \ref{tab:sjoert}.  This conversion is necessary to use the $M_\bullet - \mathcal{M}_{\rm bulge}$ relation in our estimates of SMBH mass (Eq. \ref{eq:KH13}).

Given the progenitor masses, we now follow the formalism of \citet{Taffoni+03} to estimate $T_{\rm DF}$, the dynamical friction inspiral time that sets the delay between time of merger and the onset of the TDE rate enhancement.  We approximate the dark matter halos of the two progenitor galaxies as NFW profiles \citep{Navarro+97} with mass density
\begin{equation}
\rho_{\rm NFW}(r) = \frac{M_{\rm h}}{4\pi R_{\rm h}^3} \frac{\delta_{\rm c}}{c_{\rm h}x(1+c_{\rm h}x)^2}.
\end{equation}
Here $x=r/R_{\rm h}$ is a dimensionless radius, $c_{\rm h}$ is the dimensionless halo concentration parameter and
\begin{equation}
\delta_{\rm c} = \frac{c_{\rm h}^3}{\ln(1+c_{\rm h})-c_{\rm h}/(1+c_{\rm h})}.
\end{equation}
The circular speed $V_{\rm c}^2(r) = GM(r)/r$ can be computed from the mass enclosed at radius $r$,
\begin{equation}
M(r) = M_{\rm h} \frac{\ln(1+c_{\rm h}x) - c_{\rm h}x/(1+c_{\rm h}x)}{\ln(1+c_{\rm h}) -c_{\rm h}/(1+c_{\rm h})}.
\end{equation}
For a halo of given mass $M_{\rm h}$, we define the virial radius as
\begin{equation}
R_{\rm h}= \left(\frac{3M_{\rm h}}{4\pi \times 200 \bar{\rho}} \right)^{1/3},
\end{equation}
where $\bar{\rho}(z)$ is the Universe's average density at redshift $z$.  We estimate the concentration parameters of the NFW halos using empirical results from weak lensing \citep{Mandelbaum+08}:
\begin{equation}
c = (5.6\pm0.9 ) (1+z)^{-1}\left( \frac{M_{\rm h}}{10^{14} M_\odot} \right)^{-0.13\pm 0.07}.
\end{equation}
Now we can estimate the dynamical friction inspiral time $T_{\rm DF}$ of a satellite with initial orbital energy $E$ and angular momentum $J$ using Eq. \ref{eq:DF}.  Defining the ratio of halo masses to be $Q_{\rm h} = M_{\rm s}/M_{\rm h}$, we can define the dimensionless constant $A$ from Eq. \ref{eq:DF} in two different regimes.  When $Q>0.1$, the satellite galaxy does not lose significant mass from tidal stripping during its dynamical friction-driven inspiral, and the ``rigid body'' dynamical friction time can be computed with
\begin{equation}
A_{\rm rigid} = 0.6 f(c_{\rm h}) \epsilon^\alpha x_{\rm c}^{1.97} / \ln \Lambda.
\end{equation}
Here the Coulomb logarithm $\ln \Lambda = \ln(1+Q_{\rm h}^{-1})$, $x_{\rm c} = r_{\rm c}(E)/R_{\rm h}$, $r_{\rm c}(E)$ is the initial circular orbit radius, the circularity parameter $\epsilon = J/J_{\rm c}(E)$, and the circular orbit angular momentum $J_{\rm c}(E) = r_{\rm c}(E)V_{\rm c}(r_{\rm c}(E))$.  We also define the following dimensionless support functions:
\begin{align}
f(c) =& 1.6765+0.0446c \\
\alpha(x_{\rm c}, Q_{\rm h}) =& 0.475 \left(1-\tanh(10.3Q_{\rm h}^{0.33}-7.5x_{\rm c}) \right).
\end{align}

\begin{table}
\caption{Average bulge to total (B/T) corrections for different bins of total stellar mass $\mathcal{M}_\star$ in the \citet{vanVelzen17} sample of SDSS galaxies.}\label{tab:sjoert}
\centering
\begin{tabular}{ll |  }
$\log_{10}(\mathcal{M}_\star/M_\odot)$     &  B/T   \\
\hline\hline
8.225 &   0.0944  \\
8.614 &   0.1202   \\
8.985 &   0.1205  \\
9.363 &   0.1424   \\
9.785 &   0.2777  \\
10.13 &   0.4527   \\
10.49 &   0.5637  \\
10.83 &   0.6692   \\
11.17 &   0.7685   \\
\end{tabular}	
\end{table}

However, when $Q_{\rm h}<0.1$, the satellite galaxy undergoes significant tidal stripping, increasing the inspiral time substantially.  A fitting formula for the inspiral time in this ``live satellite'' regime, valid for $Q_{\rm h}<0.08$, is
\begin{align}
A_{\rm live} =& \left[0.25\left(\frac{c_{\rm h}}{c_{\rm s}} \right)^6 - 0.07 \frac{c_{\rm s}}{c_{\rm h}}+1.123 \right] \left[B(x_{\rm c})Q_{\rm h}^{0.12} + C(x_{\rm c})Q_{\rm h}^2 \right] \notag \\
& \times \left[0.4+D(x_{\rm c}, Q_{\rm h})(\epsilon - 0.2) \right].
\end{align}
The above formula has been well tested against N-body integrations for $\epsilon \ge 0.3$, and may be unreliable for more radial orbits.  It uses the following dimensionless support functions:
\begin{align}
B(x)=& -0.050+0.335x+0.328x^2 \\
C(x)=& 2.151-14.17x+27.383x^2 \\
D(x, Q_{\rm h})=& 0.9+10^8(12.84+3.04x-23.4x^2) \\
& \times \left(Q_{\rm h}-\frac{0.0077}{1-1.08x}-0.0362 \right)^6 \notag
\end{align}
Following the suggestion of \citet{Taffoni+03}, when $0.06 \le Q_{\rm h} \le 0.1$, we interpolate linearly between these two regimes\footnote{While \citet{Taffoni+03} suggest a limiting $Q_{\rm h}=0.08$, we find that this produces non-monotonic behavior for $T_{\rm DF}(Q_{\rm h})$.  We instead pick a limiting $Q_{\rm h}=0.06$ to ensure monotonicity.}.

Throughout this paper, we employ a $\Lambda$CDM cosmology with the most recent Planck calibration of cosmological parameters: $\Omega_\Lambda=0.687$, $\Omega_{\rm M}=0.313$, and $H_0=67.48~{\rm km~s^{-1}~Mpc^{-1}}$ \citep{Planck+16}.

\end{document}